\documentclass{aa}
\usepackage[dvips]{graphicx}
\usepackage{natbib}
\usepackage{txfonts}
\usepackage{color}

\begin{document}
\newcommand{\NB}[2][]{\textbf{\color{red}{#1 #2}}}

   \title{Radiative hydrodynamics simulations of red supergiant stars: I. interpretation
of interferometric observations}
\titlerunning{Radiative hydrodynamics simulations of RSGs and interferometry}

   \author{A. Chiavassa
          \inst{1,2}
          \and
          B. Plez\inst{2,4}
          \and
          E. Josselin\inst{2}
          \and
          B. Freytag\inst{3,4}
          }

   \offprints{A. Chiavassa}

  \institute{ Max-Planck-Institut f\"{u}r Astrophysik, Karl-Schwarzschild-Str. 1, Postfach 1317, DÐ85741 Garching b. M\"{u}nchen, Germany\\
              \email{chiavass@mpa-garching.mpg.de}
            \and
            GRAAL, Universit\'{e} de Montpellier II - IPM, CNRS, Place Eug\'{e}ne Bataillon
	34095 Montpellier Cedex 05, France
	         \and
	          Centre de Recherche Astrophysique de Lyon,
          UMR 5574: CNRS, Universit\'e de Lyon,
          \'Ecole Normale Sup\'erieure de Lyon,
          46 all\'ee d'Itali.e., F-69364 Lyon Cedex 07, France
          \and
	         Department of Physics and Astronomy,
          Division of Astronomy and Space Physics,
          Uppsala University,
          Box 515, S-75120 Uppsala,
          Sweden
                         }

   \date{Received; accepted }

  \abstract
   {It has been suggested that convection in Red Supergiant (RSG) stars gives rise 
   to large-scale granules causing observable surface inhomogeneities. This
   convection is also extremely vigorous, and suspected to be one of the causes of 
   mass-loss in RSGs. It must thus be understood in details. Evidence has been 
   accumulated that there are asymmetries in the photospheres of RSGs, but detailed
   studies of granulation are still lacking.
    Interferometric observations offer an exciting possibility to tackle this question, but
    they are still often interpreted using smooth symmetrical
limb-darkened intensity distributions, or very simple spotted ad hoc models.}
   {We explore the impact of the granulation on visibility curves and closure phases using the radiative transfer code OPTIM3D. We simultaneously assess how 3D simulations of convection in RSG with CO$^5$BOLD can
be tested against these observations.}
   {We use 3D radiative-hydrodynamics (RHD) simulations of convection to compute intensity
maps at various wavelengths and time, from which we derive interferometric visibility amplitudes and
phases. We study their behaviour with time, position angle, and wavelength, and 
compare them to observations of the RSG $\alpha$~Ori.
}
   {We provide average limb-darkening coefficients for RSGs. We detail the prospects
for the detection and characterization of granulation (contrast, size) on RSGs.
 We demonstrate that our RHD simulations provide an excellent fit to existing interferometric 
 observation of $\alpha$~Ori, contrary to limb darkened disks. This confirms the existence of
 large convective cells on the surface of Betelgeuse.}
   {}

    \keywords{stars: supergiants --
                stars: atmospheres --
                hydrodynamics --
                radiative transfer --
                techniques: interferometric 
             }

   \maketitle
%

\section{Introduction}

Massive stars with masses between roughly 10 and 25~M$_{\odot}$ spend some time as
red supergiant (RSG) stars being the largest stars in the universe. They have effective temperatures, $T_{\rm eff}$, ranging from
     3\,450 to 4\,100\,K, luminosities of 20\,000 to 300\,000\,L$_\odot$
     and radii up to 1\,500\,R$_\odot$ \citep{2005ApJ...628..973L}. Their luminosities place them among the brightest
stars, visible to very large distances. There is however a number of open issues. They shed 
large amounts of mass back to the interstellar medium, but 
their mass-loss mechanism is unidentified, although Alfv\'en and acoustic waves  have 
been proposed \citep{1984ApJ...284..238H, 1989A&A...209..198P, 1997A&A...325..709C}, as well as 
acoustic waves and radiation pressure on molecules \citep{2007A&A...469..671J}.
 Their chemical composition is largely unknown, despite the work of 
 e.g. \cite{2000ApJ...530..307C}, and \cite{2007ApJ...669.1011C}, due to difficulties in 
 analysing their spectra with broad, 
asymmetric lines with variations suspected to stem from a convection pattern 
consisting of large granules and (super-)sonic velocities 
\citep{2007A&A...469..671J,2008AJ....135.1450G}. 
Progress has been made recently, with their $T_{\rm eff}$-scale being 
revised both at solar and Magellanic Clouds metallicities
using 1D hydrostatic models  
\citep{2005ApJ...628..973L,2006ApJ...645.1102L,2007ApJ...660..301M,2007ApJ...667..202L}. 
Although these MARCS models \citep{2008A&A...486..951G} give a good
fit of the optical spectra allowing the derivation of $T_{\rm eff}$ and 
reddening, problems remain. There is a blue-UV excess in many of the observed
 spectra, that may point to scattering by circumstellar dust, or to an
 insufficiency in the models. There is also a mismatch in the IR colours,
 that could be due to atmospheric temperature inhomogeneities  
characteristic of convection \citep{2006ApJ...645.1102L}.

In recent years, hydrodynamical modeling of convection in RSGs has lagged behind 
that of solar type stars due to the necessity to include the whole star in the 
simulation box. \cite{2002AN....323..213F} have succeeded in doing such numerical 
simulations of a typical RSG. We have thus engaged in an important effort to 
improve our understanding 
and description of RSGs using detailed numerical simulations and a large 
set of observational material. 

This paper is the first in this series and it is aimed to explore the 
granulation pattern of RSGs and its impact on interferometric observations.

\section{3D radiative transfer in Radiative-hydrodynamics simulation}

\subsection{3D hydrodynamical simulations with CO$^5$BOLD}\label{modelsect}

The numerical simulations employed in this work have been obtained using CO$^5$BOLD
\citep{2002AN....323..213F,
Freytag2008A&A...483..571F}
in the \emph{star-in-a-box} configuration:
the computational domain is a cube, and the grid is equidistant in all directions.
All six faces of the cube use the same open boundary conditions
for material flows and emergent radiation.
In addition, there is an "inner boundary condition":
in a small spherical region in the center of the cube
a source term to the internal energy provides the stellar luminosity
and a drag force brakes dipolar flows through it.
Otherwise, the hydrodynamics and the radiative transfer scheme
do not notice the core and integrate right through it.
Radiation transport is strictly in LTE.
The grey Rosseland mean opacity is a function of gas pressure and temperature.
The necessary values are found by interpolation in a table which has been merged
at around 12\,000\,K from high-temperature OPAL data
\citep{1992ApJ...397..717I} and low-temperature PHOENIX data
\citep{1997ApJ...483..390H} by Hans-G{\"u}nter Ludwig.
Some more technical information can be found in
\cite{Freytag2008A&A...483..571F},
the CO$^5$BOLD Online User Manual
(www.astro.uu.se/\textasciitilde bf/co5bold$\_$main.html),
and in a forthcoming paper by Freytag (2009).

\begin{figure}[h!]
   \centering
    \begin{tabular}{c}
      \includegraphics[width=1.0\hsize]{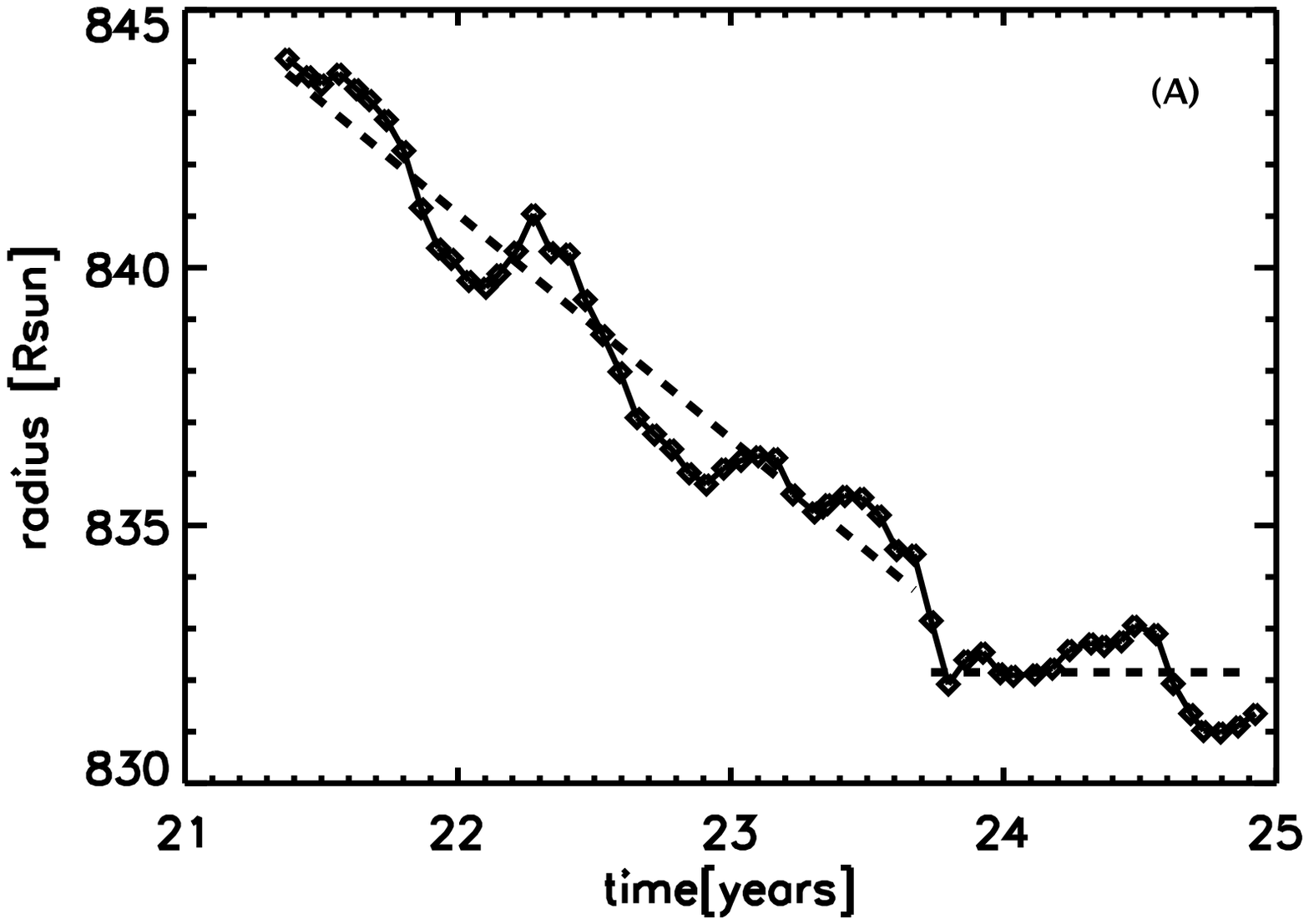} \\
  \includegraphics[width=1.0\hsize]{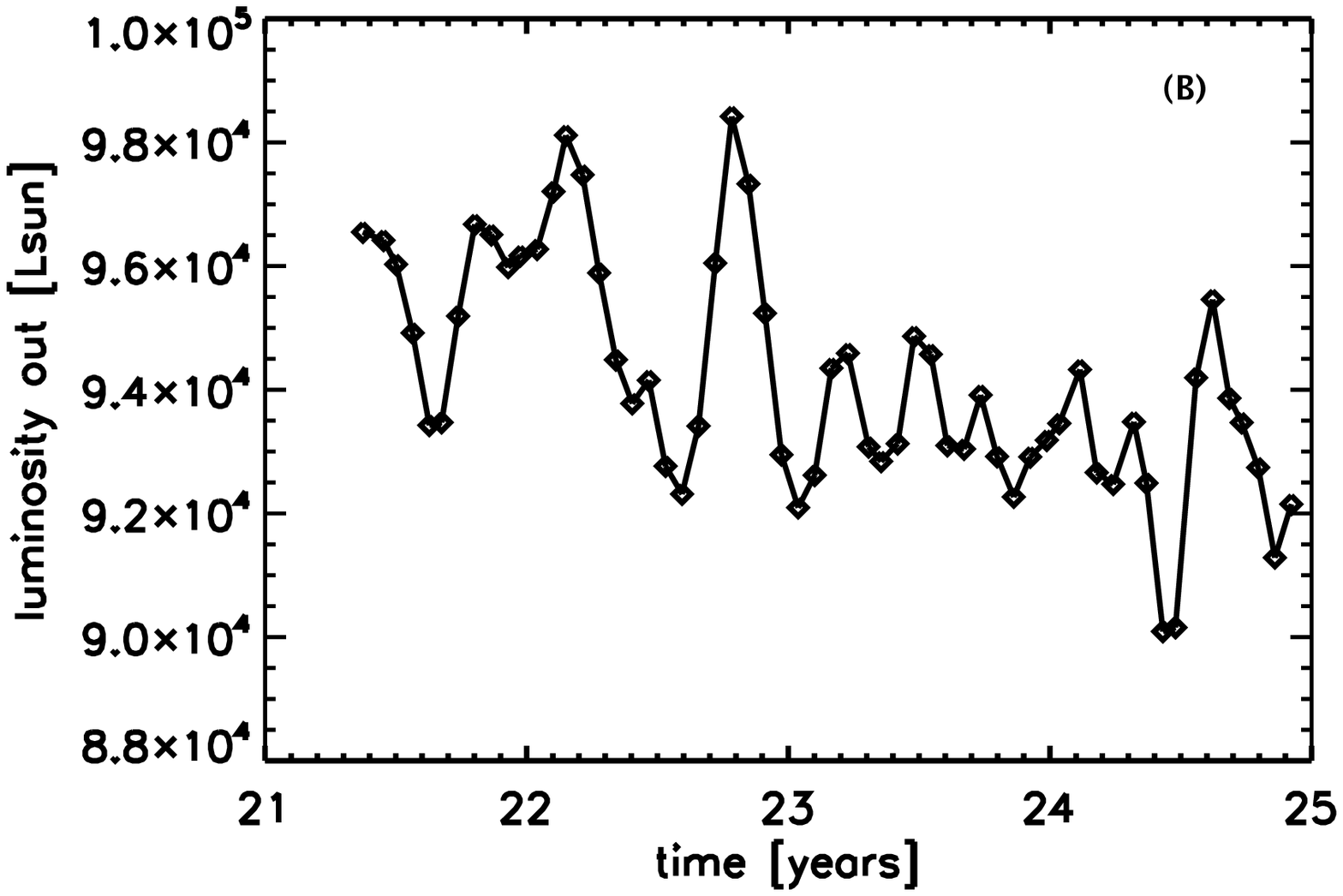} \\
      \includegraphics[width=1.0\hsize]{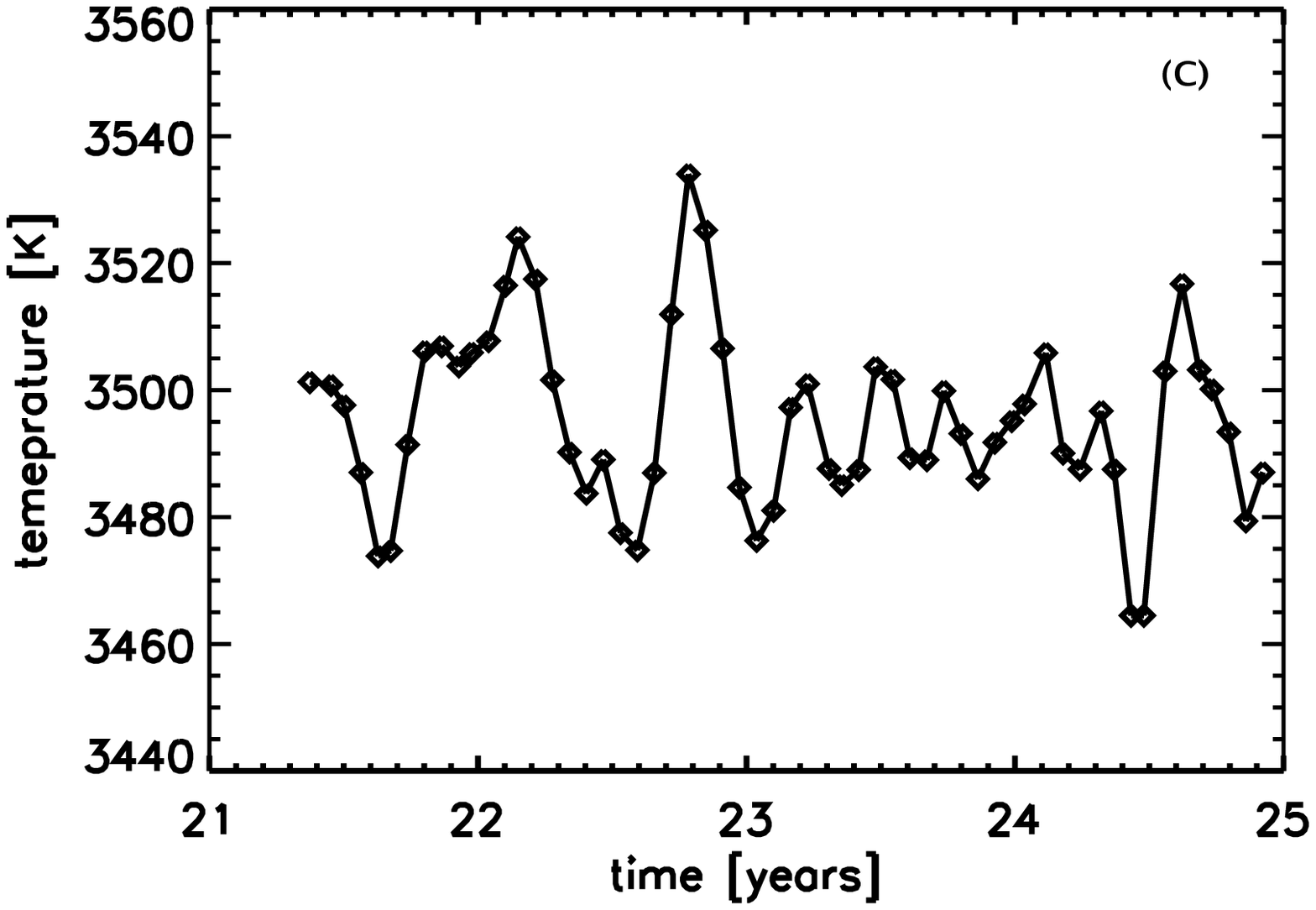} 
 \end{tabular}
      \caption{Radius (Panel A), luminosity (Panel B) and temperature (Panel C) as 
      a function of time for the simulation used in this work. The radius is fitted with the law: 
      $R(t)=936.9 - 4.3\times t$ for $t\leq 23.8$~yrs years and $R=832$~R$_{\odot}$ for $t>23.8$~yrs.
           }
        \label{quantities_vs_time}
   \end{figure}

The 12~M$_{\odot}$ model we use in this paper (st35gm03n07) is
the result of intensive calculations which have led to about 7.5 years
of simulated stellar time. It has a numerical resolution of 8.6~R$_{\odot}$ in
a cube of 235$^3$ grid points.
The model  parameters are a luminosity of $L=93000\pm 1300~{\rm L}_{\odot}$,
an effective temperature of $T_{\rm{eff}}=3490\pm 13~K$,
a radius of $R=832\pm0.7~{\rm R}_{\odot}$,
and followingly a surface gravity ${\rm log}~g=-0.337\pm0.001$.
These values are
averages over spherical shells and over time (over the last year), and
the errors are one sigma fluctuations with respect to the average over time.
We define the stellar radius, $R$, and the effective temperature,  $T_{\rm{eff}}$, as follows.
First, we compute the average temperature and luminosity over spherical shells, $T(r)$,
and $L(r)$. We then search the radius $R$ for which 
${L(R)}/(4\pi R^2)=\sigma T^4(R) \label{eq_radius_co5bold}$, where
$\sigma$ is the Stefan-Boltzmann constant. 
The effective temperature is then $T_{\rm{eff}}=T(R)$.
Fig.~\ref{quantities_vs_time} shows the value of the radius, 
temperature and luminosity over the last 3.5 years.
The radius drifts by about $-0.5\%$ per year and seems to have 
stabilised to $R=832~{\rm R}_{\odot}$
in the last year. $T_{\rm eff}$ fluctuates by $\pm 1\%$ over the
whole sequence, with a constant average. 
The luminosity fluctuations are of the order of $\pm 4\%$, reflecting 
the temperature variations, with a 
decrease of about $1\%$ per year in the first years, reflecting the 
radius decrease. These drifts indicate 
that the simulation has not completely converged in the first years. 
In this work we consider the whole 3.5 year sequence, despite the small 
radius drift, in order to have better statistics. 
The preceding 4 years of the simulation are not considered here, since they show
larger drifts. The interferometric observables derived in this work are not sensitive 
to the drift of the parameters.

This is our "best" RHD simulation so far because it has stellar parameters
closest to real RSGs (e.g., 3650~K for $\alpha$~Ori,
\citealp{2005ApJ...628..973L}). New simulations with different stellar
parameters are in progress and they will be analyzed in a forthcoming paper.

\subsection{Radiative transfer code: OPTIM3D}\label{optim3dsect}

We have developed a 3D pure LTE radiative transfer code, OPTIM3D to generate synthetic spectra and 
intensity maps from snapshots of the 3D hydrodynamical simulations, taking into account the Doppler 
shifts caused by the convective motions. The radiation transfer is calculated in detail using pre-tabulated 
extinction coefficients generated with the MARCS code \citep{2008A&A...486..951G}. These tables are 
functions of temperature, density and wavelength, and were computed with the solar composition 
of \citet{2006CoAst.147...76A}. The tables include the same extensive atomic and molecular data as 
the MARCS models. They were constructed with no micro-turbulence broadening and the temperature and 
density distribution is optimized to cover the values encountered in the outer layers of the RHD simulations. 
The wavelength resolution is $R=\lambda/\Delta\lambda=500\,000$ and we checked that this resolution is sufficient to ensure an 
accurate calculation of broadened line profiles of RSGs even after interpolation of the opacity at the 
Doppler shifted wavelengths.

The monochromatic intensity emerging towards the observer at a given position on the simulation can be computed by integrating
the source function along a ray perpendicular to a face of the cube, at that position. In LTE it reads:
\begin{equation}
{I}_{\lambda}\left(0\right)=\int^{\tau_\lambda}_0 S_{\lambda}\left({t_\lambda}\right)e^{-{t_\lambda}}d{t_\lambda}\label{optim_equation}
\end{equation}
where $I_{\lambda}$ is the intensity, $t_{\lambda}$ is the  optical depth along the ray increasing inwards, 
$\tau_\lambda$ is the maximum optical depth reached along the line-of-sight, and 
 $S_{\lambda}=B_{\lambda}\left(T\right)$, the Planck function at the temperature $T$, is the source function.
A Gauss-Laguerre quadrature of order $n$ can be performed to evaluate the integral, Eq.~(\ref{optim_equation}),
when $\tau_\lambda\rightarrow\infty$. 
This method is much faster than a detailed integration along the discretized ray, as it uses only
the value of the source function at $n$ depth points weighted with $n$ predetermined weights.
This method is reliable as long as
the source function is sufficiently smooth along the optical depth scale, and is well known at the 
quadrature points. This is not always the case in our simulations where the optical depth scale may jump 
by large amounts between 2 successive cells, e.g. from $\tau=1$ to $\tau=300$ for extreme cases. 
The source function must then
be interpolated at intermediate optical depths, and the result is largely dependent on the way this
 interpolation is performed. Note however that this is also the case for a detailed integration, where 
 too large jumps in the source function or optical depth scale will cause uncertainties in 
 the resulting intensity. We checked for differences between a Gauss-Laguerre quadrature and a detailed summation
of the contributions from all cells, with different kinds of interpolations, and found differences in 
intensities emerging from a single ray
of less than $10\%$ on average, with some extreme
cases reaching more than $100\%$ due to a particularly illconditioning
of the source function. 
The average differences being in an acceptable range, we therefore rely on the Gauss-Laguerre quadrature, 
with a linear interpolation of the source function on
the logarithmic $\tau$-scale. 
The quadrature points and weights we use are listed in Tab.~\ref{tab_gauss} \citep{abramo}. 
For the rays where the optical depth does not reach high enough values, 
we carry a detailed
summation of the contribution from all cells along the ray.

In practice, once the input simulation is read, OPTIM3D interpolates the opacity tables in temperature and logarithmic density 
for all the simulation grid points using a bi-linear interpolation. The interpolation coefficients are computed only once, and stored. 
Bi-linear interpolation has been preferred to spline interpolation because: (i) spline is significantly more time consuming, 
(ii) and comparisons with other codes do not show great improvements using splines (see below).
Then, the logarithmic extinction coefficient is linearly interpolated at each Doppler-shifted wavelength in each cell along the ray, 
and  the optical depth scale along the ray is calculated. 
Eq.~(\ref{optim_equation}) is then integrated, giving the intensity emerging towards the observer at that wavelength and position.
 This calculation is performed for every line-of-sight perpendicular to the face of the computational box, and for all the required 
 wavelengths.
  \begin{table}
      \caption[]{Gauss-Laguerre quadrature weights for n=10.}
         \label{tab_gauss}
     $$ 
         \begin{array}{p{0.5\linewidth}l}
              \hline
                        \hline
            \noalign{\smallskip}
             {\rm abscissa} & {\rm weight} \\ 
            \noalign{\smallskip}
            \hline
            \noalign{\smallskip}
 0.137793470540 & 3.08441115765E-01 \\ 
     0.729454549503 & 4.01119929155E-01 \\ 
    1.808342901740 & 2.18068287612E-01 \\ 
    3.401433697855 & 6.20874560987E-02 \\ 
    5.552496140064 & 9.50151697518E-03 \\ 
    8.330152746764 & 7.53008388588E-04 \\ 
    11.843785837900 & 2.82592334960E-05 \\ 
    16.279257831378 & 4.24931398496E-07 \\ 
    21.996585811981 & 1.83956482398E-09 \\ 
    29.920697012274 & 9.91182721961E-13 \\
            \noalign{\smallskip}
            \hline
         \end{array}
     $$ 
   \end{table}
   
Comparisons with existing codes were carried out.
The spectral synthesis code Turbospectrum (\citealp{1993ApJ...418..812P}, 
\citealp{1998A&A...330.1109A}, and further improvements by Plez) was used with
 one-dimensional MARCS models, where the source fonction is very well
 sampled on the $\tau$-scale. OPTIM3D computations made with bi-linear interpolation
 deviate by  less than $5\%$, and the deviation decreases to 0.2$\%$ with
 spline interpolation.
We also checked OPTIM3D against Linfor3D (\citealp{2007A&A...473L..37C} for the Non-LTE version, and 
http://www.aip.de/\textasciitilde mst/Linfor3D/linfor\_3D\_manual.pdf
for the LTE version) using 3D CO$^5$BOLD local models. We compared synthetic spectra computed for three artificial 
iron lines (with increasing strength) centered at a laboratory wavelength of 5500\ \AA \ and using the same abundances. 
The discrepancy between the results of the codes is less than 3$\%$ and it becomes even less than 0.2$\%$ when a 
spline interpolation of the opacity tables is used in OPTIM3D (with a significant increase of the CPU time). 
Finally, comparisons were made with the spectral line formation code used by, e.g., \cite{2000A&A...359..755A} for 3D 
local convection simulations carried out with the code by \cite{1998ApJ...499..914S} for giant stars (\citealp{2007A&A...469..687C}). 
The tests have been carried out on the [OI] line at 6300.3\,\AA\ and various Fe\,I and Fe\,II lines around 5000\,\AA. 
The discrepancies between the resulting synthetic spectra are less than 2$\%$, and become even less than 0.6$\%$ when a 
spline interpolation of the opacity tables is used in OPTIM3D. Thus, the interpolation is the main source of error. 
In conclusion, if only a few lines are computed for, e.g., accurate abundance determinations, Linfor3D 
or the Asplund code are better because 
they mostly avoid interpolations into opacity tables. The code used by Asplund performs bi-cubic interpolations of the continuum
opacity and of the individual number densities, whereas we interpolate the total opacity from all lines contributing at a given
wavelength, which is in principle less accurate. So, when a large  wavelength range must 
be calculated taking into account many  molecular and atomic lines simultaneously, OPTIM3D is a better, faster choice, that still  provides
a result accurate to a few percents.
\section{Simulated images in the H and K bands: giant convective cells}\label{radiusSect}

\begin{figure}
   \centering
   \includegraphics[width=1\hsize]{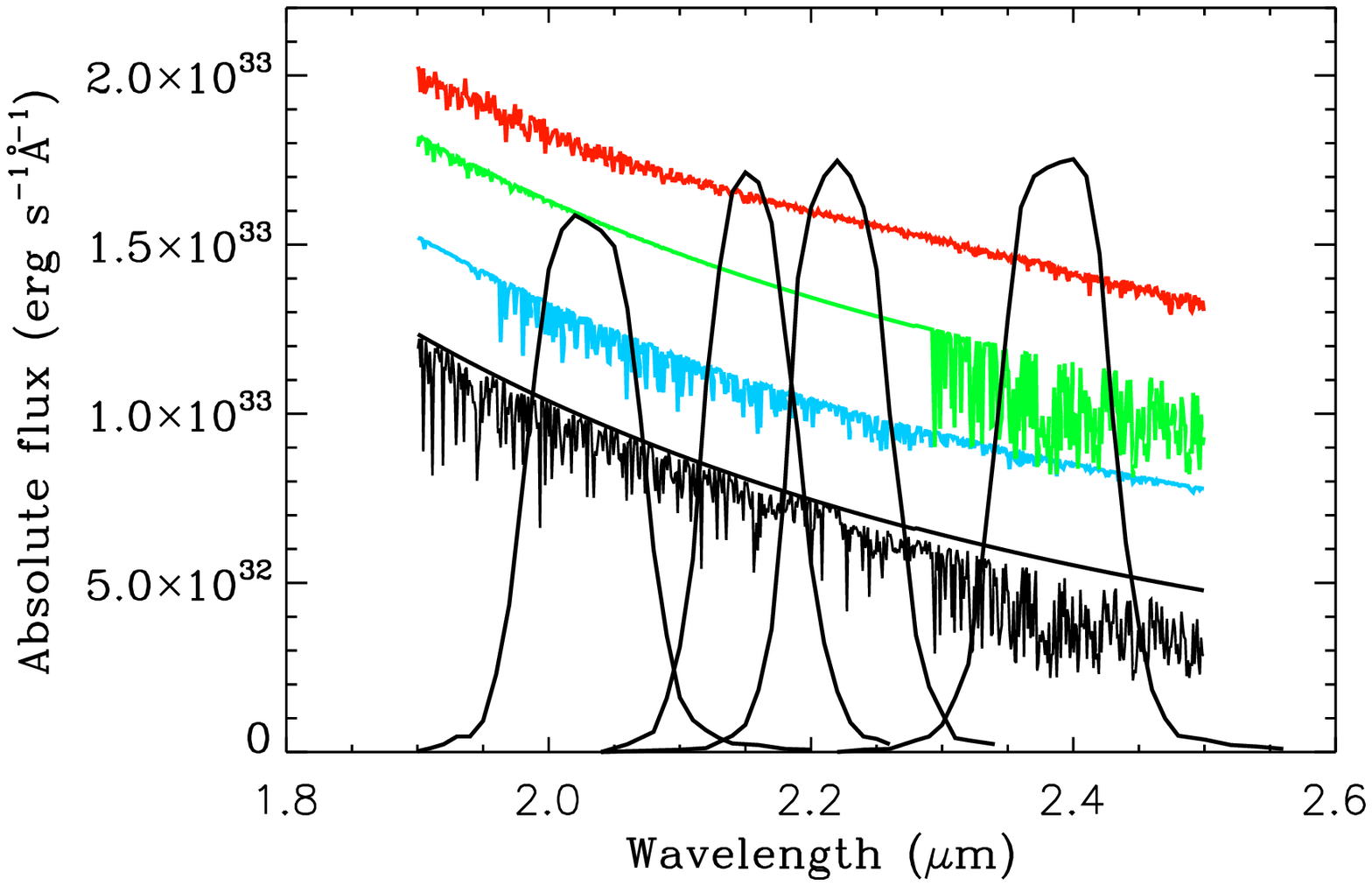}
      \caption{The transmission curves of the 4 narrow band filters mounted on the FLUOR instrument at 
      IOTA together with the K band synthetic spectrum of a snapshot of the simulation and the 
      corresponding continuum (bottom black curve). From the top, the spectra computed with only 
      H$_2$O (red), only CO (green), and only CN (blue) are shown
      with an offset of respectively $0.9\times 10^{33}$, $0.6\times 10^{33}$ and 
      $0.3\times 10^{33}$~erg\,cm$^{-2}$\,s$^{-1}$\,\AA$^{-1}$. 
           }
         \label{filterFLUOR}
   \end{figure}

\begin{figure}
   \centering
   \includegraphics[width=1\hsize]{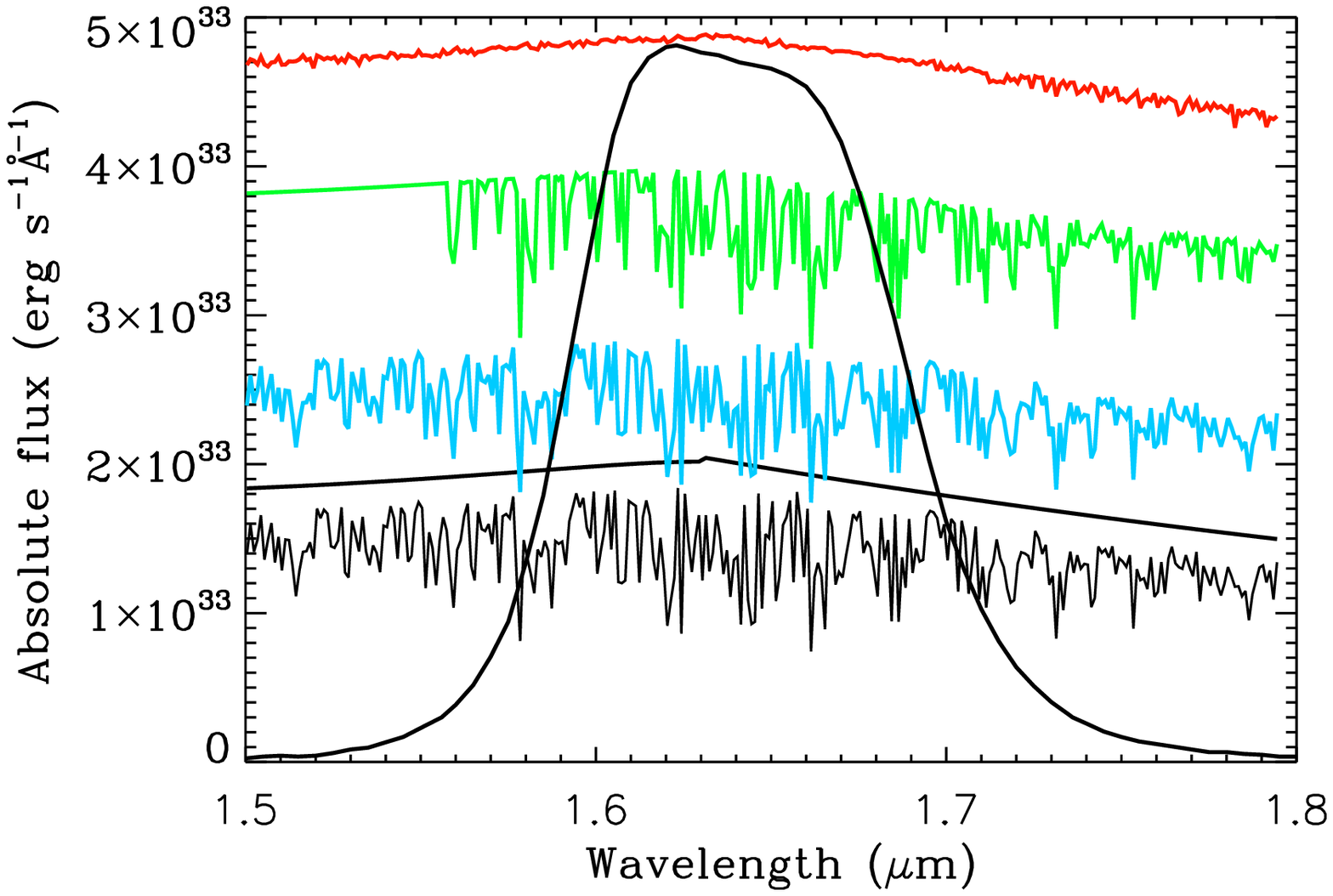}
      \caption{The transmission curve of the filter mounted on IONIC at IOTA together with the H band synthetic spectra
      computed as in Fig.~\ref{filterFLUOR}. From the top, the offset of the spectra is $3\times 10^{33}$, 
      $2\times 10^{33}$, and $1\times 10^{33}$~erg\,cm$^{-2}$\,s$^{-1}$\,\AA$^{-1}$.
      }
         \label{filterIONIC}
   \end{figure}

In this work, we analyze the properties of the simulations in the H and K bands where many interferometric 
observations have been done, and existing interferometers, e.g. VLTI/AMBER, operate routinely. We calculated  
intensity maps for a series of snapshots about $23$ days apart covering 3.5 years of the model described above.
 We use the transmission curve of the four K band filters mounted on FLUOR (Fiber Linked Unit for Optical 
 Recombination; \citealp{1998SPIE.3350..856C}), and of the H band filter mounted on IONIC (Integrated Optics 
 Near-infrared Interferometric Camera; \citealp{2003SPIE.4838.1099B}) at the IOTA interferometer (\citealp{2003SPIE.4838...45T}). 
The K band filters (Fig.~\ref{filterFLUOR}) are: K203 (with a central wavelength of $2.03$ $\mu$m), 
K215 ($2.15$ $\mu$m), K222 ($2.2$ $\mu$m),
and K239 ($2.39$ $\mu$m). 
The H band filter has a central wavelength of $1.64$ $\mu$m (Fig.~\ref{filterIONIC}). 
The resulting intensities reported in this work are normalized to the filter transmission as: 
$\frac{\int I_{\lambda} T\left(\lambda\right)d\lambda}{\int T\left(\lambda\right)d\lambda}$ where 
$I_\lambda$ is the intensity and $T\left(\lambda\right)$ is the transmission curve of the filter 
at a certain wavelength. The intensity maps are showed after applying a median [3x3] smoothing (see Section~\ref{numerical_smooth}).

\begin{figure*}
   \centering
   \begin{tabular}{c}
   \includegraphics[width=1\hsize]{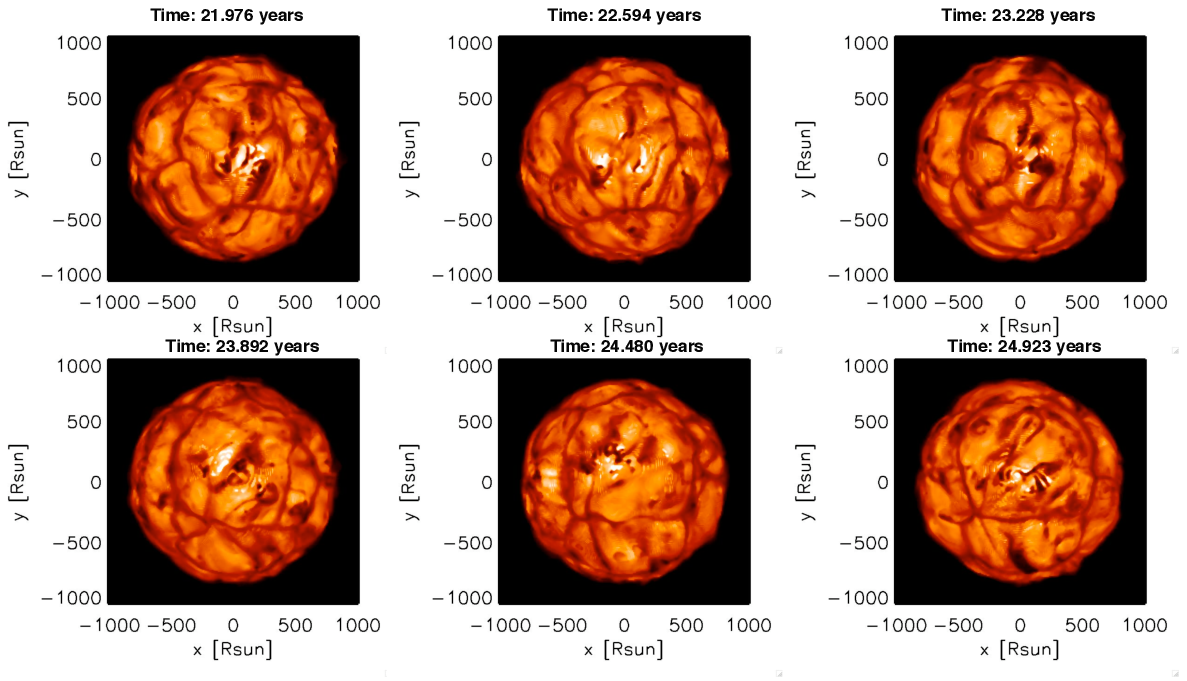}\\
   \includegraphics[width=1\hsize]{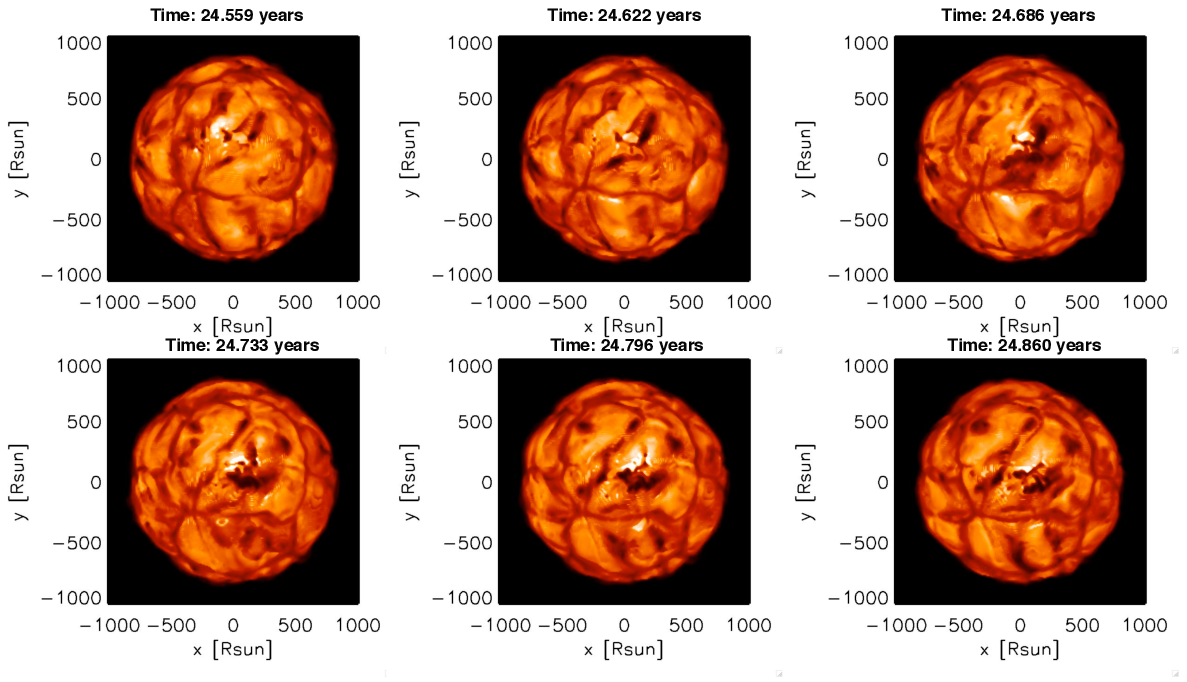}
   \end{tabular}
      \caption{\emph{Top 6 panels: }maps of the intensity in the IONIC filter (linear scale with a  range of 
       [0;$2.5\times10^5$]\,erg\,cm$^{-2}$\,s$^{-1}$\,{\AA}$^{-1}$). The different  panels correspond to  
       snapshots separated by 230 days ($\sim3.5$ years covered).
      \emph{Bottom 6 panels:} successive snapshots separated by 23 days ($\sim140$ days covered).
      }
         \label{Hband_images}
   \end{figure*}

It can be seen from our simulations (see Fig.~\ref{Hband_images})
that the surface of the stellar model is covered by few large convective cells of a size of about
 400 to 500~R$_{\odot}$ that evolve
on a time-scale of years.
These cells have strong downdrafts that can penetrate
down to the stellar core (Freytag et al. 2002, and 2009, in preparation).
Near the surface, there are short-lived (a few months to one year)
small-scale (50 to 100~R$_{\odot}$) granules (bottom panels of Fig.~\ref{Hband_images}).
\cite{1997svlt.work..316F} found a relation between the mean horizontal size 
of convective granules $x_{\rm{gran}}$ and the atmospheric pressure scale-height defined as 
$H_{p0}=\frac{kT_{\rm{eff}}}{g\mu m_{\rm{H}}}$ for GK dwarfs and subgiants. It is unclear if 
such a relation can be extrapolated 
to 3D simulations of RSGs. Using it we find $x_{\rm{gran}}/R_{\star}=10\times H_{p0}/R_{\star}=0.1$, 
for parameters appropriate for a RSG atmosphere dominated by gas pressure. 
Obviously, this leads to a size much smaller than what can be seen in Fig.~\ref{Hband_images}. 
\cite{1997svlt.work..316F} found that a value of $x_{\rm{gran}}/H_{p0} = 10$ would fit 2D simulations for 
GK dwarfs and subgiants, but they show also that A-type and F-type stars lie above the curve indicating 
that they have larger granules. These stars have large turbulent pressure that may dominate over 
the gas pressure in turbulent convective layers.
Following \cite{2008A&A...486..951G}, we write $P_{\rm{turb}}=\beta \rho v_{\rm{turb}}^2$, where $v_{\rm{turb}}$ is the 
turbulent velocity, $\rho$ is the gas density, and $\beta$ is a parameter close to one, whose 
value depends on the anisotropy of the velocity field. A better way to express 
$H_{p0}$ is thus $H_{p0}= \frac{kT_{\rm{eff}}}{g\mu m_{\rm{H}}}\left(1+\beta\gamma\left(\frac{v_{\rm{turb}}}{c_s}\right)^2\right)$, 
where $\gamma$ is the adiabatic exponent, and 
$c_s$ the sound speed. If $v_{\rm{turb}}$ is only a factor 2 larger than $c_{s}$, $H_{p0}$ is increased 
by a factor of about 5. This is the case for our RSG simulation where $P_{\rm{turb}}/P_{\rm{gas}}\sim2$ at the surface, $R_*$, as
determined in Sect.~\ref{modelsect}. 
This gives then $x_{\rm{gran}}/R_{\star}$=0.5, extrapolating 
\cite{1997svlt.work..316F} formula. This is more consistent with the large granules visible on intensity maps
 in Fig.~\ref{Hband_images}. There are further mechanisms that might influence the size of the granules: 
 (i) in RSGs, most of the downdrafts will not grow fast enough to reach
    any significant depth before they are swept into the existing deep and strong
    downdrafts enhancing the strength of neighboring downdrafts; 
    (ii) radiative effects and smoothing of small fluctuations can cause an enhancement of
  growth time for small downdrafts while the granule crossing time is short 
  due to large horizontal velocities; (iii) sphericity effects, see for example \cite{Freytag2007sf2a.conf..481F},
   and \cite{2007AN....328.1054S}; (iv) \cite{1997svlt.work..316F} use the effective temperature and the
  pressure scale height at the bottom of the photosphere as reference, however, 
  also layers below the photosphere can matter; (v) numerical resolution (or lack of it) could matter.

\section{Intensity profiles}\label{LDsect}
The simulated RSG atmospheres appear very irregular, permeated with structure and dynamics. 
The surface inhomogeneities and their temporal evolution induce strong variations of the emerging spectra, and 
intensity profiles. In this section we analyse the average centre-to-limb intensity profiles, and their time
variations.
\subsection{Surface inhomogeneities and temporal evolution}\label{surface_inomo}
\begin{figure*}
   \centering
    \begin{tabular}{cc}
      \includegraphics[width=0.5\hsize]{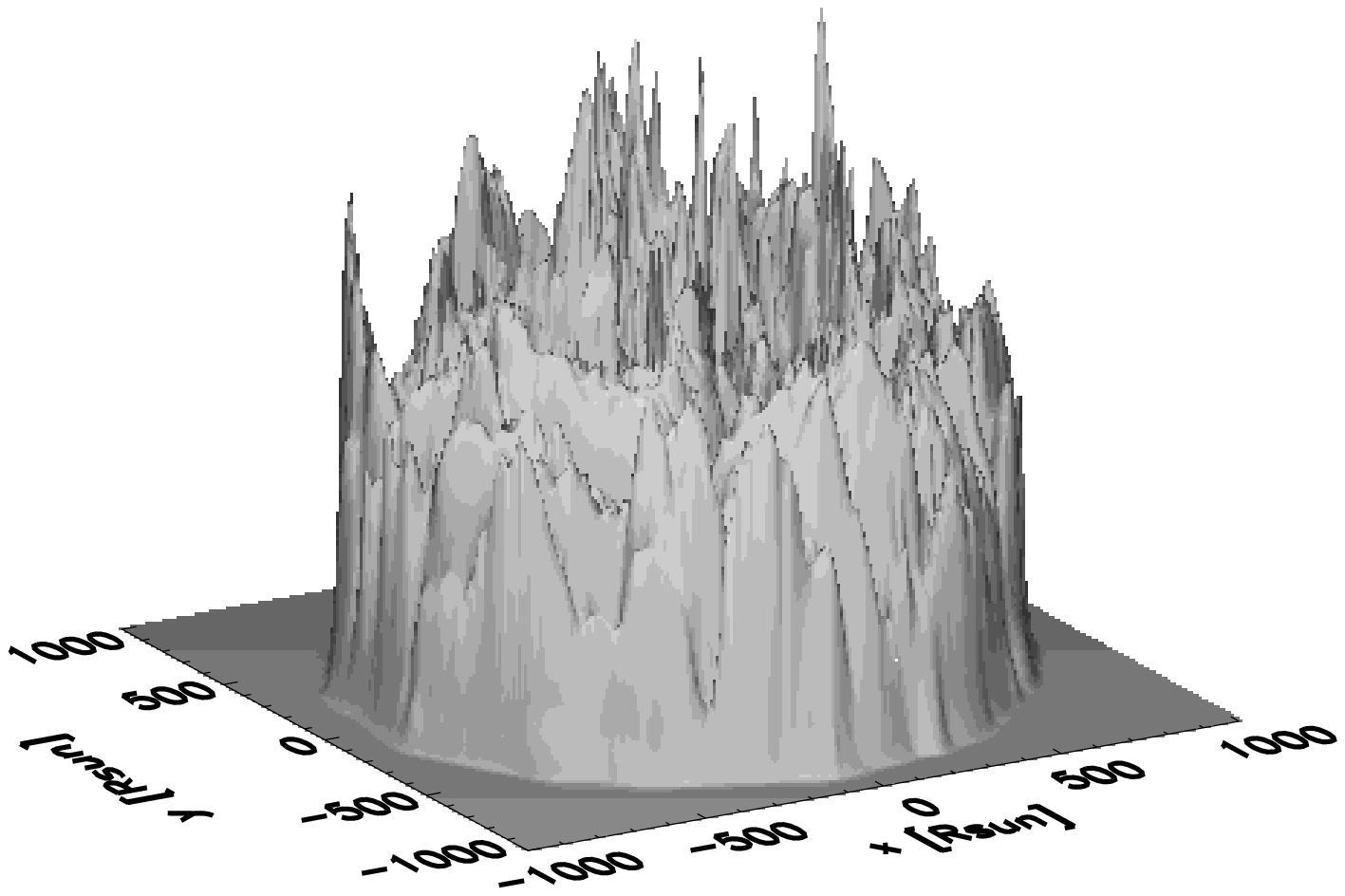} 
      \includegraphics[width=0.5\hsize]{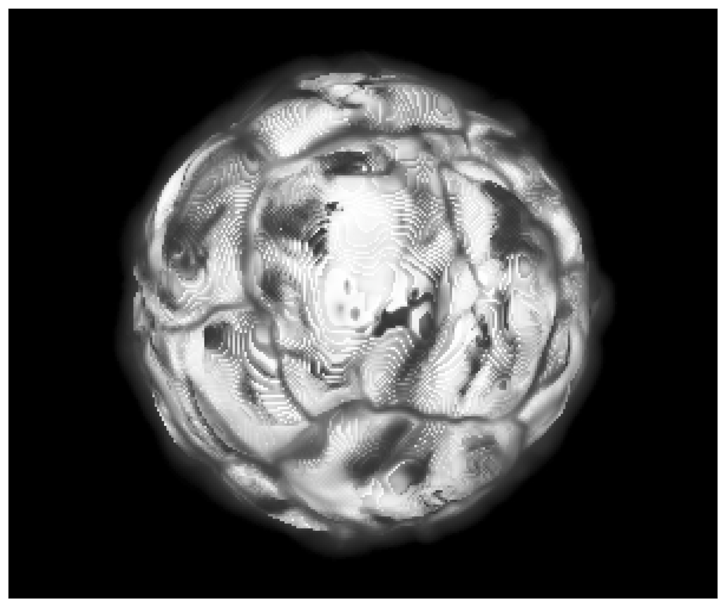}\\
      \includegraphics[width=0.5\hsize]{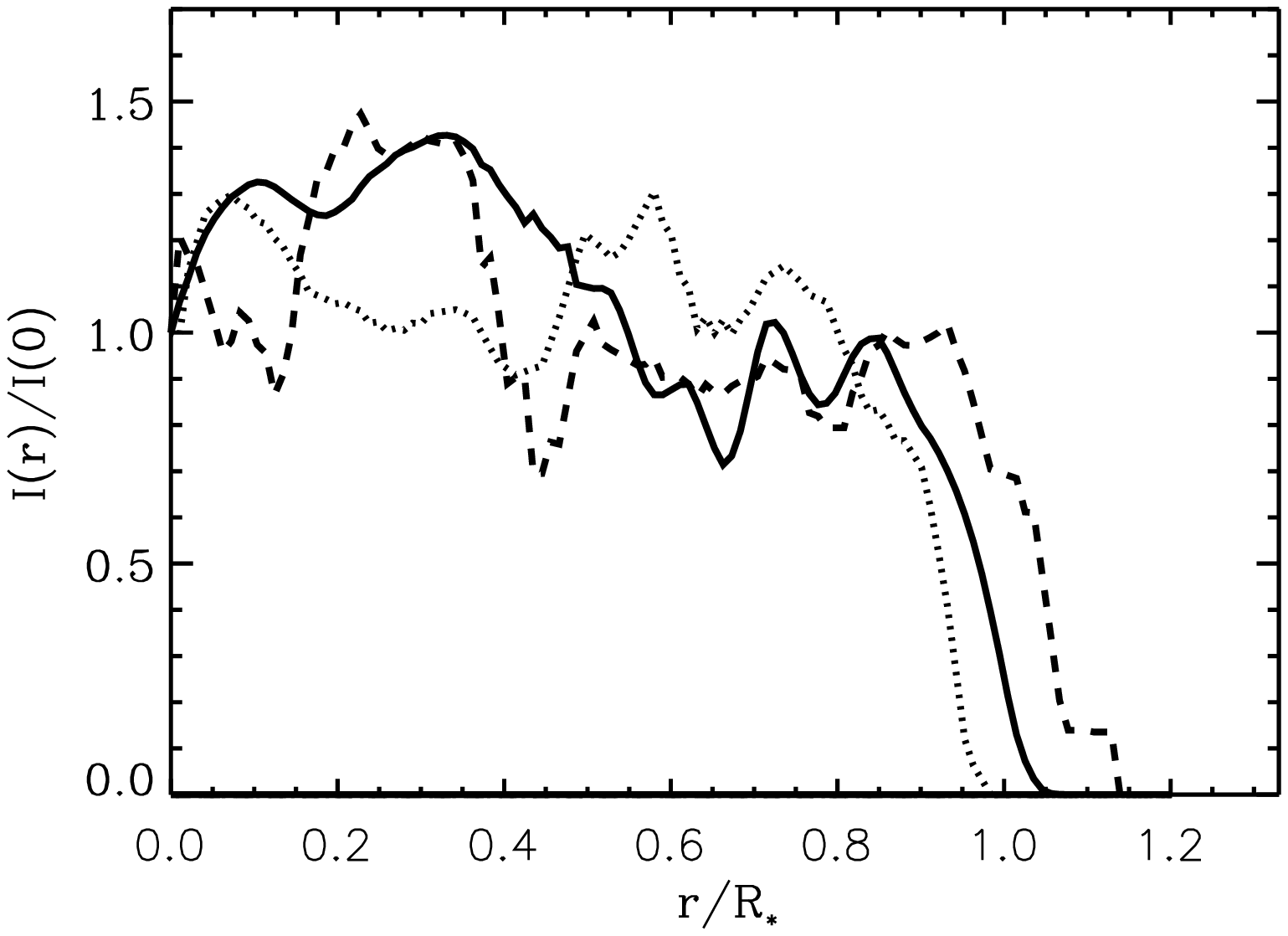}
      \includegraphics[width=0.5\hsize]{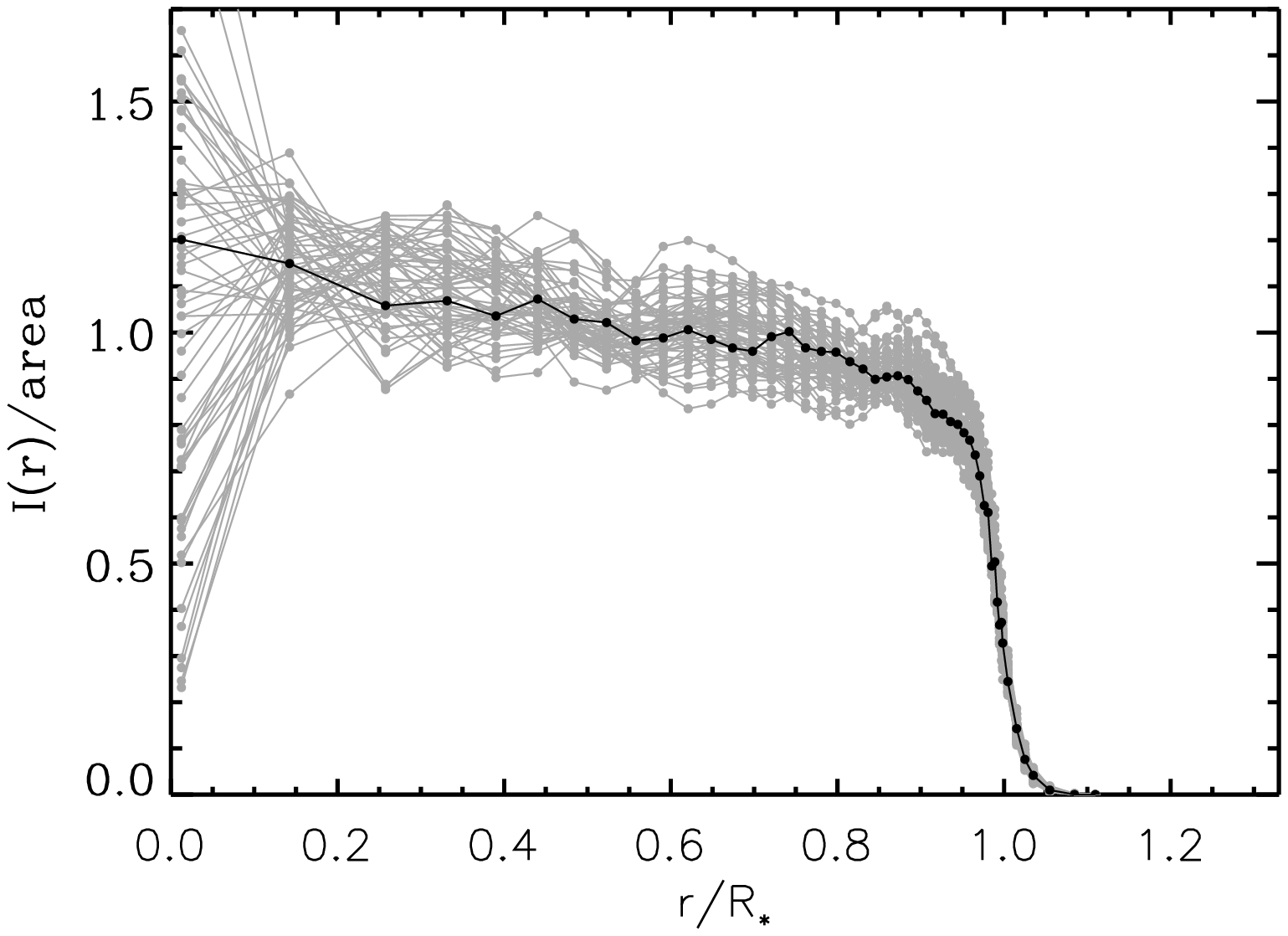}\\
 \end{tabular}
      \caption{
      \emph{Top left panel: }three-dimensional image of a snapshot from Fig.~\ref{Hband_images}. 
      \emph{Top right panel:} Intensity map of the same snapshot represented using the histogram equalization 
      algorithm in order to underline the thin bright patches due to the undersampling of the $\tau$ scale. 
      \emph{Bottom left panel:} intensity profiles for three position angles of the same snapshot. The numerical box 
      edge is at impact parameter $r/R_{\star}\sim1.3$. The intensity is normalized to the intensity at disk center. 
      \emph{Bottom right panel:} radially averaged intensity profiles for all the snapshots of the simulation (grey); 
      one snapshot of the simulation is emphasized with a solid black line. The intensity is normalized to the area 
      subtended by the curve, area$=\int_0^{1.3} I\left(r/R_{\star}\right) dr/R_{\star}$.
}
       \label{intensity_radial}
     \end{figure*}

The top left panel of Fig.~\ref{intensity_radial} shows a three-dimensional image representation of the intensity 
emerging from one face of a snapshot of the simulation in the H band. 
The K band appearance is similar. This image shows very sharp intensity peaks two to three pixels wide. This is also noticeable 
in the top right panel of the Figure as small bright (up to $40\%$ brighter than the surrounding points) patches. 
These patches result from the ill-conditioning of the source function, due to the lack of spatial resolution
around $\tau_\lambda=1$ along some lines of sight where the source function may have a large jump
(see Sect.~\ref{optim3dsect}). 
Attempts have been made to solve this problem through the interpolation of the source function and opacity inside CO$^5$BOLD, 
but they caused numerical instabilities. The unique solution is to increase the number of grid points, and that 
necessitates larger and faster computers.

Radial intensity profiles within a given snapshot show large variations with position angle of
their radial extension of about $10\%$ (see bottom left panel of Fig.~\ref{intensity_radial}). 
The variation with time of the intensity profiles are of the same order of magnitude (10$\%$, see
bottom right panel of Figure).

\subsection{The limb darkening law}
Despite the large azimuthal variations of the intensity profiles,
and their temporal variations,
it is interesting to derive radially averaged intensity profiles for each snapshot.
These may be be used, e.g., as a first approximation to interpret interferometric observations, 
in replacement of limb-darkening (LD) laws computed from hydrostatic models \citep{2000A&A...363.1081C}.
Bottom right panel of Fig.~\ref{intensity_radial}
shows all the radially averaged intensity profiles obtained from the simulation.

 We use a LD law of the form :
\begin{equation}\label{claret_law}
\frac{I(\mu)}{I(1)}=\sum_{k=0}^3 a_k\left(1-\mu\right)^k
\end{equation}
where $I(\mu)$ is the intensity, $a_k$ are the LD coefficients and $\mu=cos\theta$ with $\theta$ the 
angle between the line of sight and the radial direction. $\mu$ is related to the impact parameter 
$r/R_{\star}$ through $r/R_{\star}=\sqrt{1-\mu^2}$, where R$_{\star}$ is the stellar radius 
determined as in Sect. \ref{modelsect}.
The average intensity  profiles were constructed using rings regularly spaced in $\mu$ 
 for $\mu\le1$ (i.e., $r/R_{\star}\le1$), and adding a few points for $r>R_{\star}$  up to 
 the numerical box limit. The standard deviation of 
the average intensity, $\sigma_{I\left(\mu\right)}$, was computed within each ring. 
There is a small tail at $r>R_{\star}$ that gives a minor contribution to the total flux 
(less than 1$\%$, see bottom right panel of Fig.~\ref{intensity_radial}), and
cannot be fitted with Eq.~\ref{claret_law}.
We fitted the radially average profiles of all the snapshots of the simulation 
 (57 profiles 23 days apart covering 3.5 years). The fit was
weighted by $1/\sigma_{I\left(\mu\right)}$ in order 
to decrease the importance of central points with poor statistics. 
 The fit was first made on $I(\mu)/I_{\rm{norm}}$, with 
$I_{\rm{norm}}$ the intensity normalized to the area subtended by the curve: 
$I_{\rm{norm}}=\frac{I\left(r/R_\star\right)}{\int_0^{1.3} I\left(r/R_\star\right) dr/R_\star}$.
This was done in order to diminish the impact of intensity fluctuations between snapshots
on the fitting coefficients.

\begin{figure}
   \centering
       \includegraphics[width=1.0\hsize]{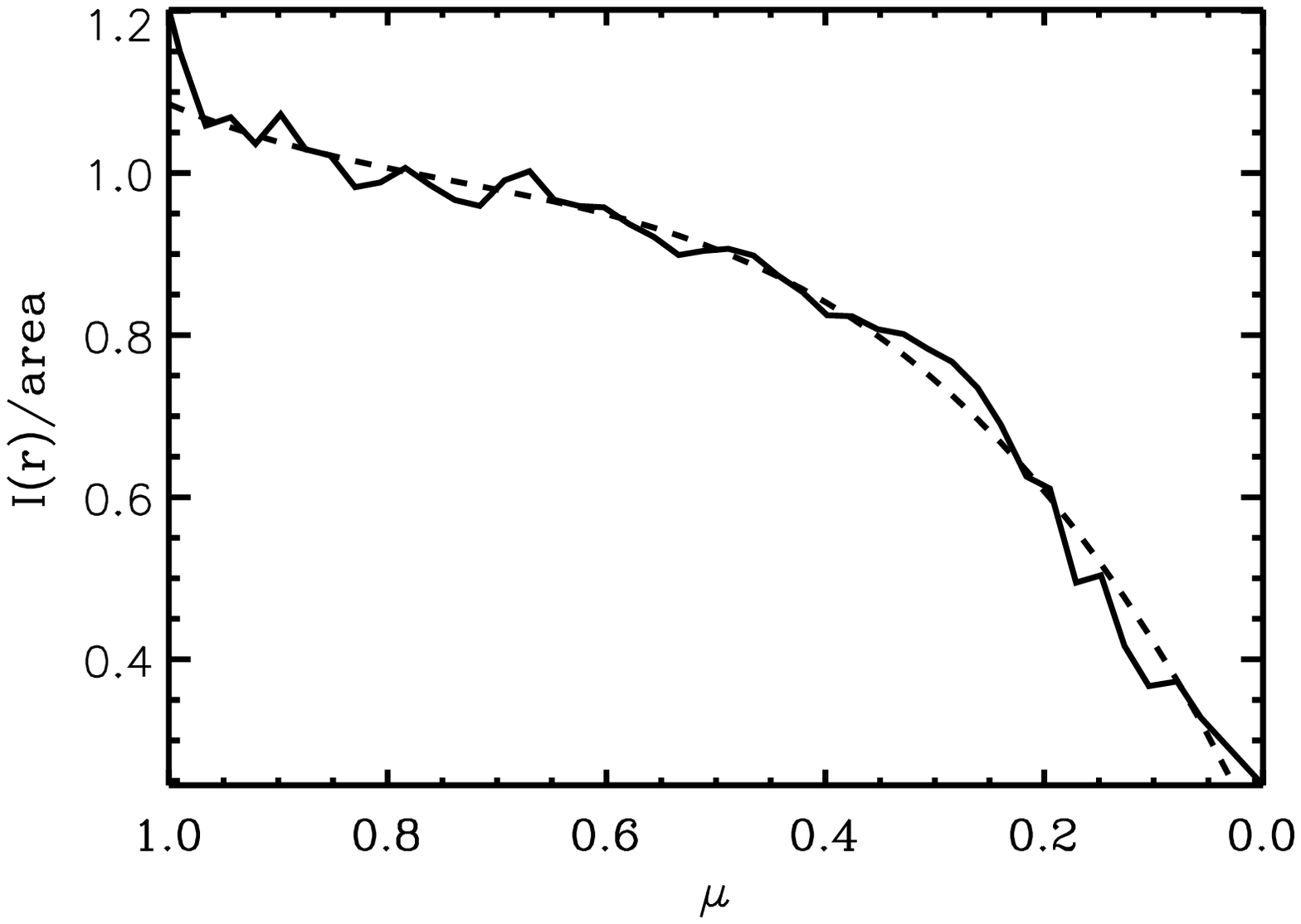}
      \caption{Example of a LD fit (dashed line) using the LD law described in the text for the radially averaged
      intensity profile (solid line) emphasized in Fig.~\ref{intensity_radial} (bottom right panel). 
      The intensity is normalized to the area subtended by the curve. This best fit has a $\chi^2=0.02$. 
            }
         \label{limbfit}
   \end{figure}

\begin{figure}
   \centering
       \includegraphics[width=1.0\hsize]{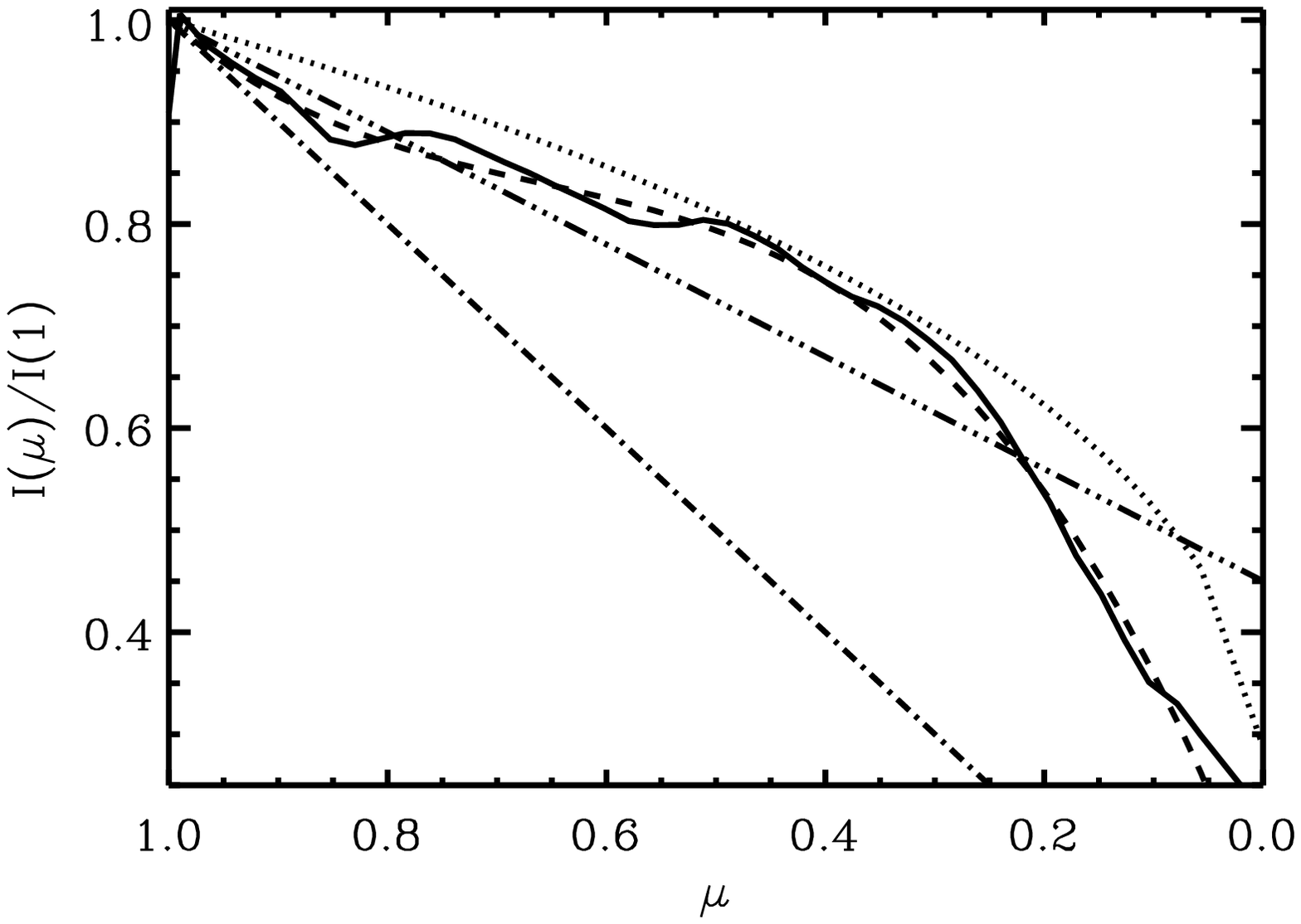}
      \caption{The time averaged H-band radial
      intensity profile of our simulation (solid line), and the fit of
      Tab.~\ref{table1} (dashed line). A fully LD (dash-dotted line), a
      partially LD (triple dot-dashed line), and a LD fit from 
      \citet[][ dotted line]{2000A&A...363.1081C} for comparable RSG parameters are plotted for
      comparison.
            }
         \label{LD-fits}
   \end{figure}

In Tab.~\ref{table1}, we give the values of the four LD coefficients averaged over all 3.5 years, and renormalized to disk center, 
for the IONIC H-band filter, and for the K222 filter (because the sensitivity of the FLUOR instrument is 
always better in the continuum 
 than in molecular bands 
 \citealp{2004A&A...426..279P}, and it
samples the maximum transmission region of the K band).
Fig.~\ref{limbfit} shows an example of LD fit. The intensity profiles for different position 
angles for the same snapshot being very different (see Fig.~\ref{intensity_radial}, bottom left panel), 
the fitting coefficients are very scattered. The time averaged LD fits give however an indication 
of the shape of the intensity profile in the H and K bands (note that they are very similar). 
They are of course very different from simple first order LD 
laws. They also differ from LD laws calculated by \citet{2000A&A...363.1081C}
for parameters appropriate for RSGs (see Fig.~\ref{LD-fits}).
When comparing to observations
of RSGs, we recommend to use our fits. Ideally one should use
single snapshots as we do below in our analysis of Betelgeuse, 
as they may deviate from the average LD fit
by large amounts (see Tab.~\ref{table1}).

%
%
\begin{table}
\begin{minipage}[t]{\columnwidth}
\caption{Time-averaged limb-darkening coefficients for the RHD simulation (see Eq.~\ref{claret_law}. 
$\sigma$ is the standard deviation over time (57 profiles covering 3.5 years.}
\label{table1}
\centering
\renewcommand{\footnoterule}{}  
\begin{tabular}{lcccccccc}
\hline \hline
$\lambda$  & $a_{0}$   &  $\sigma$   & $a_{1}$ &  $\sigma$  &  $a_{2}$&  $\sigma$  &  $a_{3}$ &  $\sigma$ \\
($\mu$m) & ~  &  ($\%$) & ~& ($\%$) & ~&($\%$) &  ~&($\%$) \\
\hline
1.64\footnote{central wavelength of the corresponding IONIC filter}  & 1.00  & 5 &  -0.93 &  50 & 2.03  &  50  &  -1.98 & 55    \\
2.22\footnote{central wavelength of the corresponding K222-FLUOR filter} &  1.00  & 5  &  -0.85  &  50  &  2.12 & 45  & -2.13 & 45   \\
\hline
\end{tabular}
\end{minipage}
\end{table}

\section{Visibility curves and phases}\label{numerical_smooth}
\subsection{Computation}
The granulation pattern has a significant impact on interferometric visibility curves and phases. 
We try here to derive their characteristic signature.

\begin{figure} [!h]
   \centering
    \includegraphics[width=0.95\hsize]{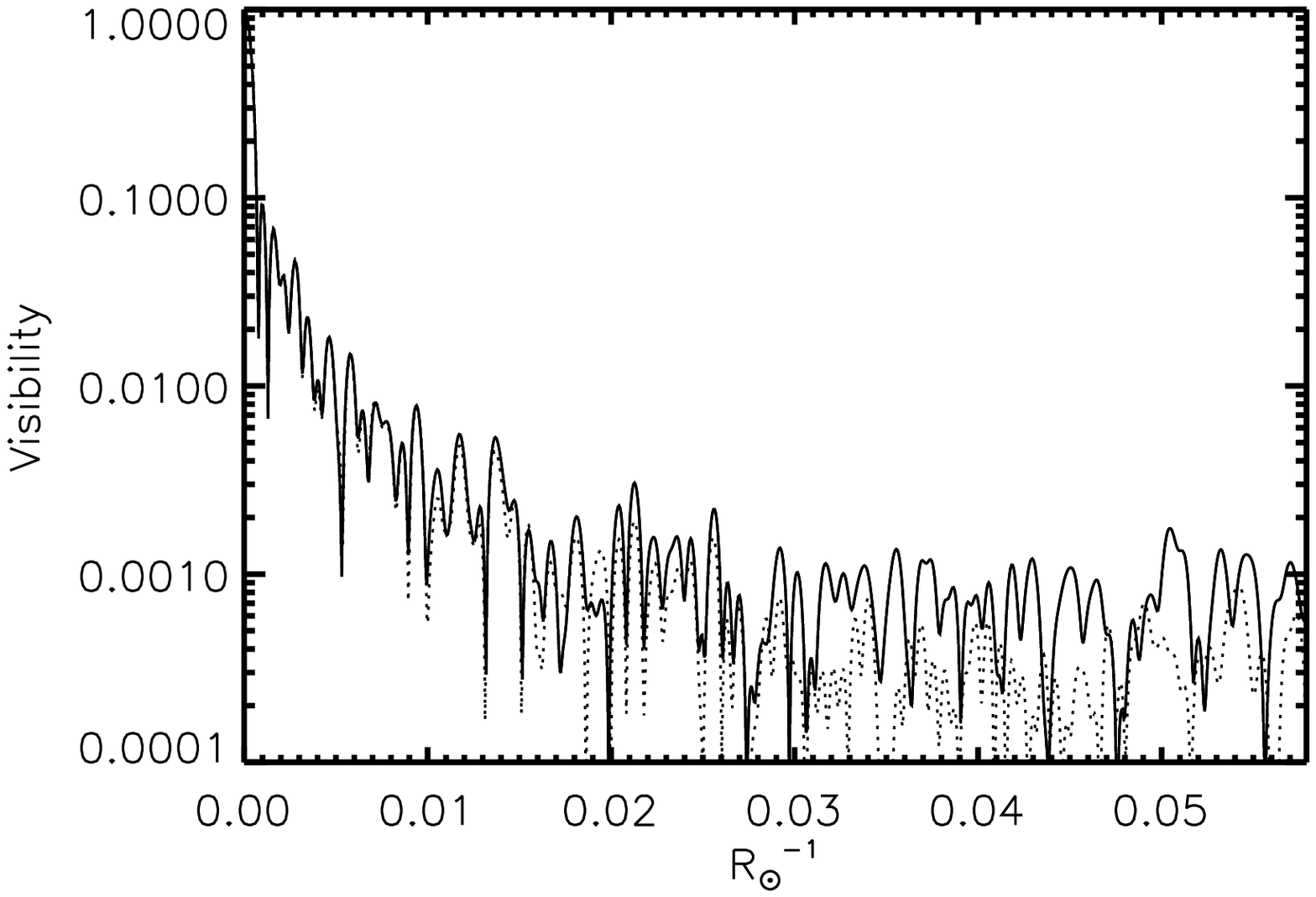}	
         \caption{The solid line is the visibility curve for the IONIC filter intensity map of 
         Fig.~\ref{intensity_radial} (top right). The dotted line is computed for the same map 
         after applying a $[3\times3]$ median smoothing.               }
         \label{histo_equ}
   \end{figure}

We compute visibilities and phases using the IDL data visualization and analysis platform. 
For each image, a discrete Fourier transform is calculated. 
In order to reduce the problems due to the finite size of the object and to avoid edge effects, the resolution 
in the Fourier plane is increased by padding the input $235\times 235$~pixels image with zeros up to a 
size of $2048\times 2048$~pixels.
The visibility $V$ is defined as the modulus $|z|$, of the complex Fourier transform,
$z = x + iy$, normalized to the modulus at the origin of the frequency plane,
$|z_0|$, 
with the phase $\tan\theta = \Im(z)/\Re(z)$.
When dealing with observations, the natural spatial frequency unit is arcsec$^{-1}$.
As we study theoretical models, we use instead R$_\odot^{-1}$ units.
The conversion factor between these is 
\begin{equation} \label{conversion}
V~[arcsec^{-1}]=V~[{\rm R}_\odot^{-1}]\cdot d~[{\rm pc}]\cdot214.9,
\end{equation}
where 214.9  is the astronomical unit expressed in solar radii, and
$d$ is the distance of the observed star.
The relation between the baseline, $B$ (in m), of an interferometer, and the spatial 
frequency $\nu$ (in arcsec$^{-1}$)
observed at a wavelength $\lambda$ (in $\mu{\rm m}$) is
$\nu=B/\lambda/0.206265$.
As our calculated images are affected by the source function jumps, we investigated 
how the visibility curves are affected by the resulting bright spikes (Fig.~\ref{intensity_radial}). 
We compare in Fig.~\ref{histo_equ} the visibility curves computed for one projected baseline 
from the raw image, and after applying a median $[3\times3]$ smoothing effectively erasing 
the artifacts. The visibility curves are affected by these artifacts mostly for frequencies greater than
 0.03~R$^{-1}_\odot$ (corresponding to 33~R$_\odot$, i.e., $\sim$4~pixels). We can therefore apply this 
 cosmetic median filter, as it will not affect the visibilities at lower frequencies, that are 
 the only ones to be observable in practice with modern interferometers.


We now study the first few lobes of the visibility curves of our simulations, and how they are 
affected by asymmetries and surface structure.

\subsection{The first lobe}\label{first_lobe_section}

\begin{figure}
   \centering
        \begin{tabular}{cc}
     \includegraphics[width=1.0\hsize]{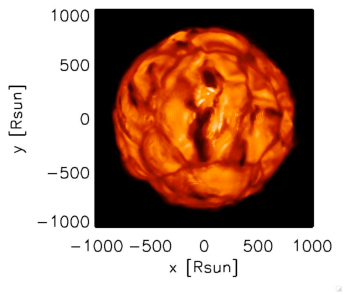}  \\
 \includegraphics[width=1.0\hsize]{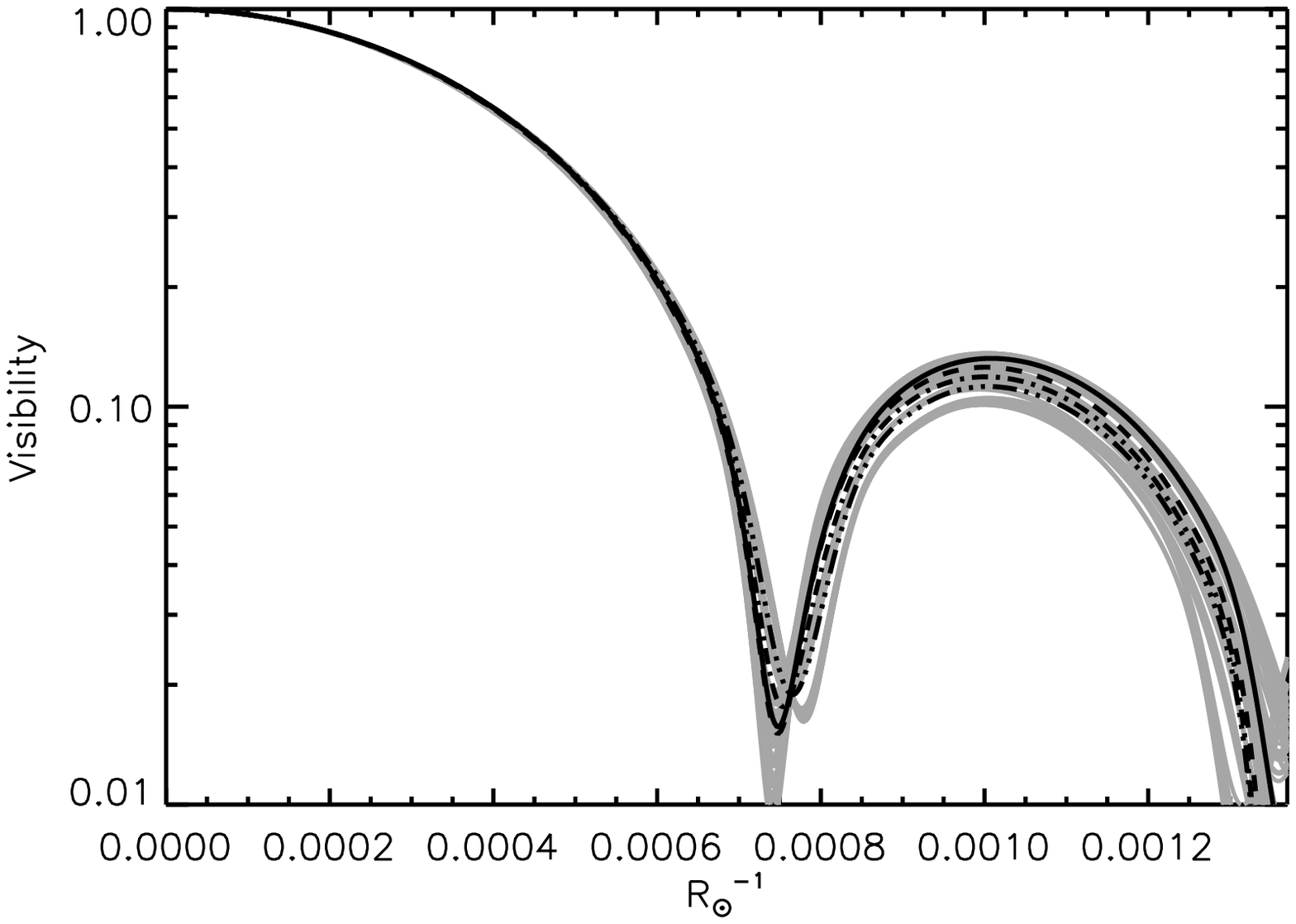} 
 \end{tabular}
      \caption{\emph{Top panel:} intensity map in the IONIC filter (the range is [0;$2.5\times10^5$]\,erg\,cm$^{-2}$\,s$^{-1}$\,{\AA}$^{-1}$). 
      \emph{Bottom panel:} visibility curves from the above snapshot computed for 36 different angles
       5$^\circ$ apart (thin grey lines). Note the logarithm visibility scale. 
       The solid black curve is a UD model, with a radius of 810~R$_\odot$. 
       The dashed black line is a partially LD disk with a radius of 822~R$_\odot$. 
       The dot-dashed line is a fully LD disk with a radius of 830~R$_\odot$.
       the triple-dot-dashed line is our average LD law (cf. Tab.~\ref{table1}) for a radius of 842~R$_\odot$.
       The stellar parameters of this snapshot are: $L = 98\,400\,{\rm L}_\odot$, 
       $R_* = 836.5\,{\rm R}_\odot$, $T_{\rm eff} = 3534$\,K and $\log~g = -0.34$.
                       }
         \label{vis_zoom1}
   \end{figure}

The first lobe of the visibility curve is mostly sensitive to the radial 
extension of the observed source. Fig.~\ref{vis_zoom1} (bottom panel) shows the visibility 
curves computed for 36 different angles from the intensity map of one snapshot in the IONIC 
filter (top panel). A dispersion of the visibility curves (thin grey lines in Fig.~\ref{vis_zoom1})
is noticeable.  This behavior is similar for all the snapshots. 
These synthetic visibilities have been compared to a uniform disk (UD) model 
(solid line in Fig.~\ref{vis_zoom1}), and with limb-darkened (LD) models. 
We use both a fully limb-darkened disk ($I_{\mu}/I_1=\mu$, dotted-dashed line in the Figure), 
and a partially limb-darkened model with $a_1=-0.5$ ($I_{\mu}/I_1=0.5+0.5\cdot\mu$, dashed line in the Figure).
The radius determined by fitting a UD disk model to the computed visibilities ranges
from 794 to 845~R$_\odot$ for the 36 angles, 
up to 5$\%$ 
 smaller than $R_{\star}$=836.5~R$_\odot$, the radius of the simulation 
 determined as described in Sect.~\ref{modelsect}.
The partially-, and fully-darkened models are respectively $\sim 2\%$, and only $\sim 1\%$ 
smaller than $R_{\star}$. 
In Fig.~\ref{vis_zoom1} is also shown the visibility amplitude resulting from our average LD fit 
of Tab.~\ref{table1}. The resulting diameter is then 842~R$_\odot$, very close to the simulation radius.
Note, that \cite{2006A&A...454..327N} also found that the UD radius is about 4 to 5$\%$ smaller than 
 the photospheric radius of their simulation of Cepheids, and that the 
 LD radius is much closer to the radius of their simulation.
Stellar diameters determined with UD or LD fits of observed first visibility lobe of RSGs will be affected
by these systematic errors. As will be shown below, observations of higher spatial frequencies
will greatly improve the knowledge of the limb-darkening, and of asymetries, thus helping in better
constraining the radius as well.

\begin{figure} 
   \centering
  \includegraphics[width=1.0\hsize]{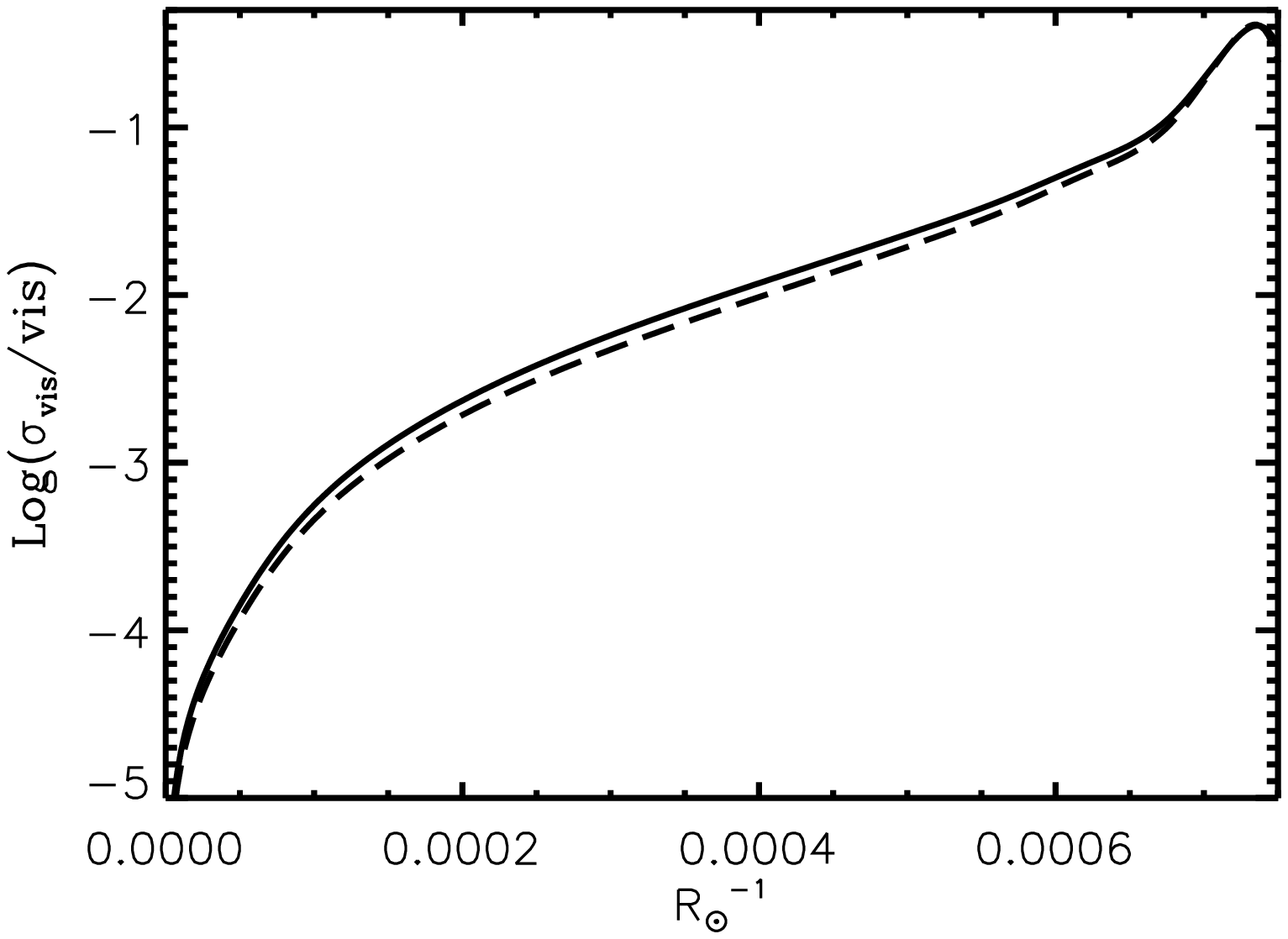} 
       \caption{standard deviation of the visibility normalized to the visibility in the first lobe.
       The solid line indicates the temporal fluctuations for one fixed angle over 3.5~years. 
       The trend is similar for the other angles. 
       The dashed curve corresponds to the angular fluctuations of the snapshot in Fig.~\ref{vis_zoom1}. }
         \label{fluct_vis_1}
   \end{figure}

It is interesting to compare the angular and temporal visibility fluctuations at one sigma, 
defined as $F=\sigma/\rm{vis}$: 
(i) the \emph{temporal evolution}, fixing one angle and following the RHD simulation for 3.5~years 
with a time-step of $\sim$23~days; 
(ii) the \emph{angular evolution}, considering a single snapshot and computing the visibilities for 
36 different angles 5$^\circ$ apart. Fig.~\ref{fluct_vis_1} shows that, in the first lobe, temporal 
and angular fluctuations have the same order of magnitude. The fluctuations are less than 
1$\%$ at frequency $\sim$0.00040~R$^{-1}_\odot$ 
(at this frequency, the visibility is greater than $50\%$), 
they are $\sim3\%$ at frequency 0.00057~R$^{-1}_\odot$, and are close to $\sim10\%$ at 0.00069~R$^{-1}_\odot$.

\subsection{The second, third and fourth lobes: signature of the convection}

\begin{figure}
\centering \includegraphics[width=1.0\hsize]{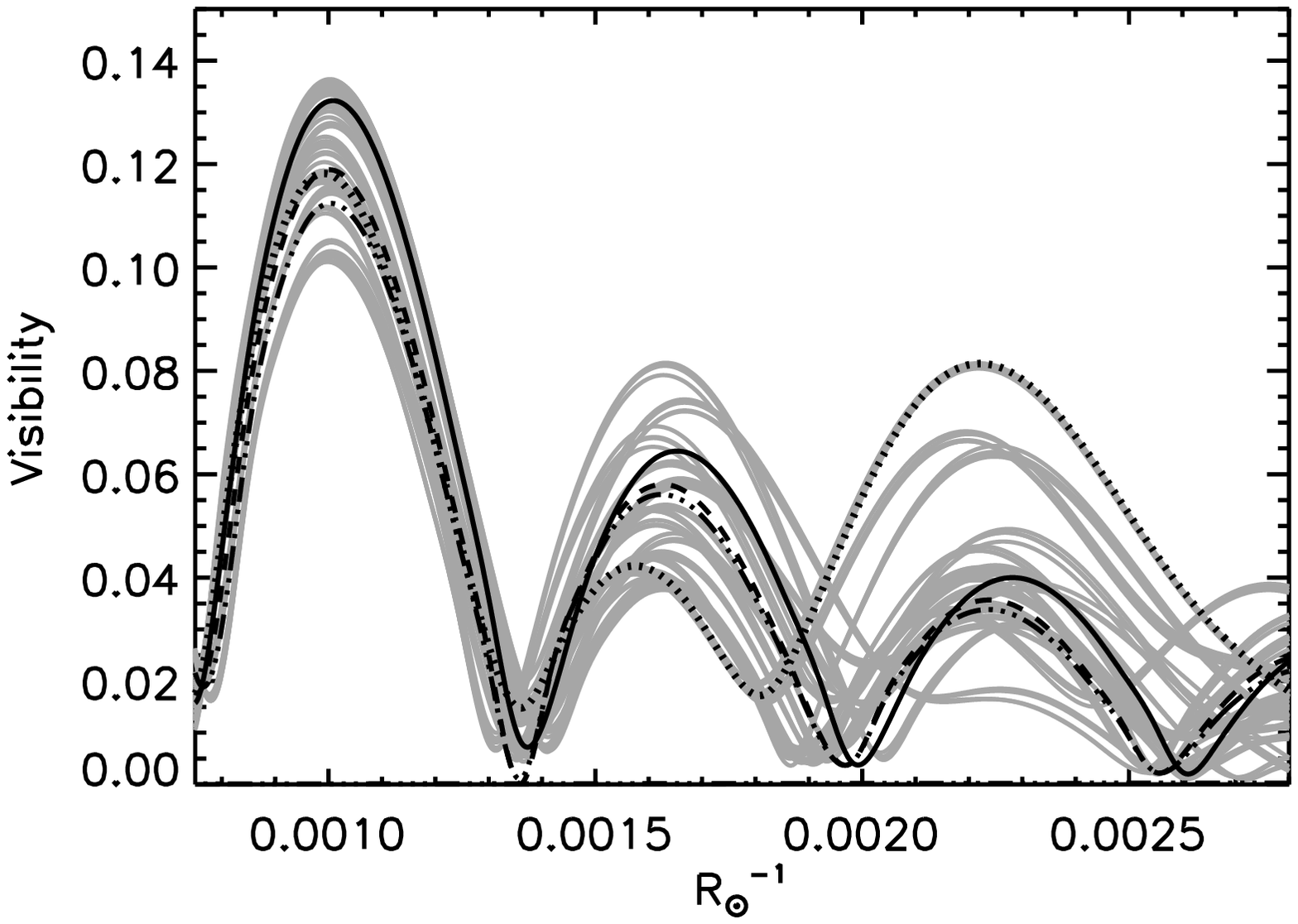}  
          \caption{Same as Fig.~\ref{vis_zoom1} for the second, third, and fourth lobes.
          In addition, the dotted line is the visibility curve for the particular azimuth
          parallel to the x-axis of the IONIC intensity map of  Fig.~\ref{vis_zoom1}.
            }
         \label{vis_zoom2}
   \end{figure}

\begin{figure} 
   \centering
 \includegraphics[width=1.0\hsize]{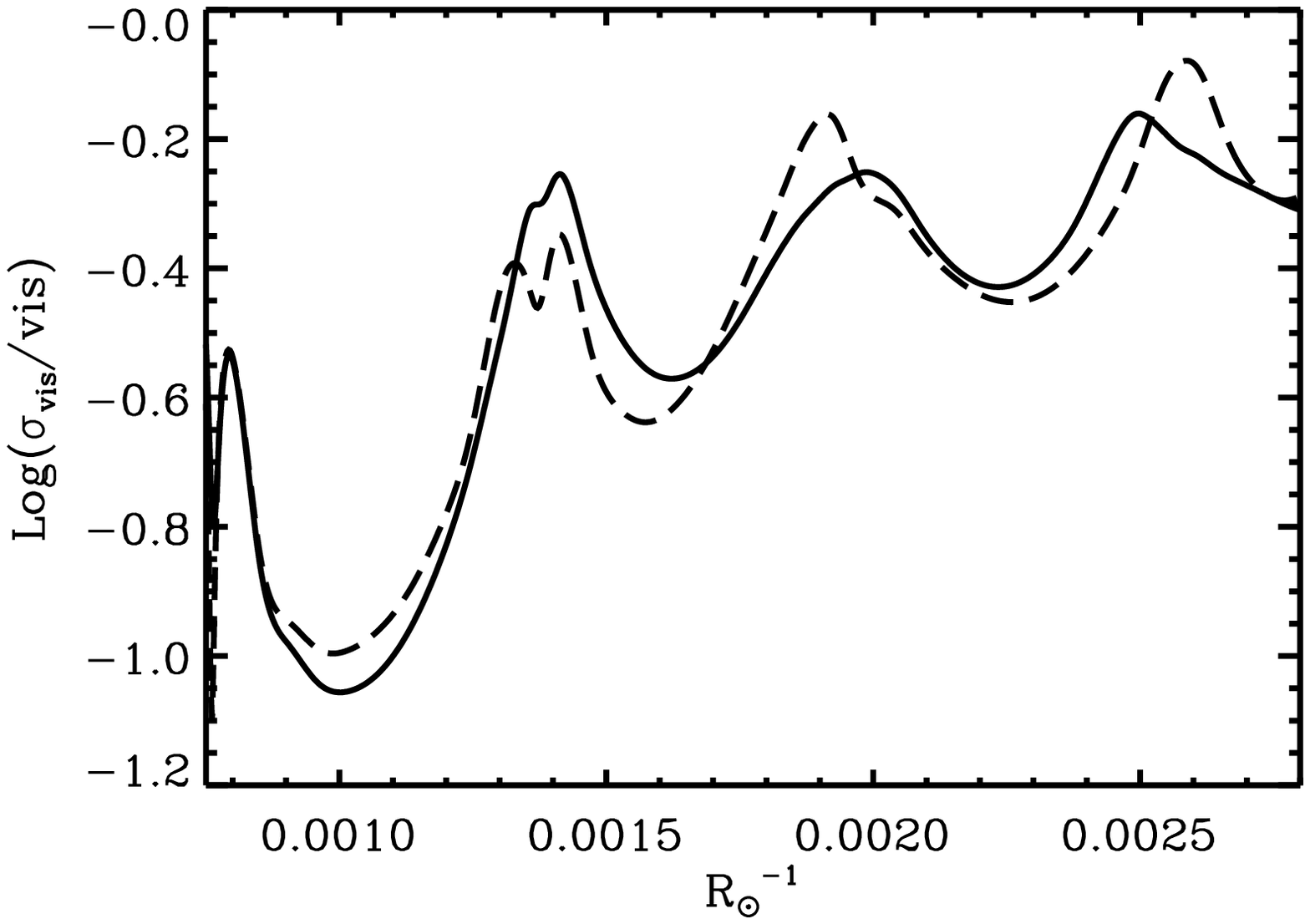} 
      \caption {Same as Fig.~\ref{fluct_vis_1} for the second, third, and fourth lobe.
      }
         \label{fluct_vis_2}
   \end{figure}

As in Sect. \ref{first_lobe_section}, we analyze the angular, and temporal visibility fluctuations 
at one sigma  with respect to the average value in the H band (IONIC filter). 
Fig.~\ref{vis_zoom2} shows an enlargement of the the second, third and fourth lobes of the visibility curves
computed for different position angles. The dispersion increases clearly with spatial frequency, and 
visibilities deviate greatly from the UD or LD cases, due
to the small scale structure on the model stellar disk.
The same is true for temporal fluctuations of the visibility at a given position angle. 
Fig.~\ref{fluct_vis_2} shows the temporal 
fluctuations of the visibilities for one fixed position angle, as well as the angular fluctuations 
for the snapshot  of Fig.~\ref{vis_zoom1}. As for the first lobe, there is no clear distinction between 
angular and temporal fluctuations. Relative fluctuations are of course large around the minima of visibility, 
where visibilities are also more difficult to measure. However, with the precision of current interferometers
(e.g., 1$\%$ for visibilities of $\sim5-10\%$ for VLTI-AMBER), it should be possible to characterize the 
granulation pattern on RSGs. This requires observing the third and the fourth lobes and not limiting the 
observation at the first and second lobes, that give only an information on the radius and LD. 
The signal to be expected in these lobes is higher than the UD or LD predictions (see dashed line in Fig.~\ref{vis_zoom2}). 
Efforts should therefore be put on observing at these frequencies.

\begin{figure*}
   \centering
    \begin{tabular}{c}
       \includegraphics[width=0.5\hsize]{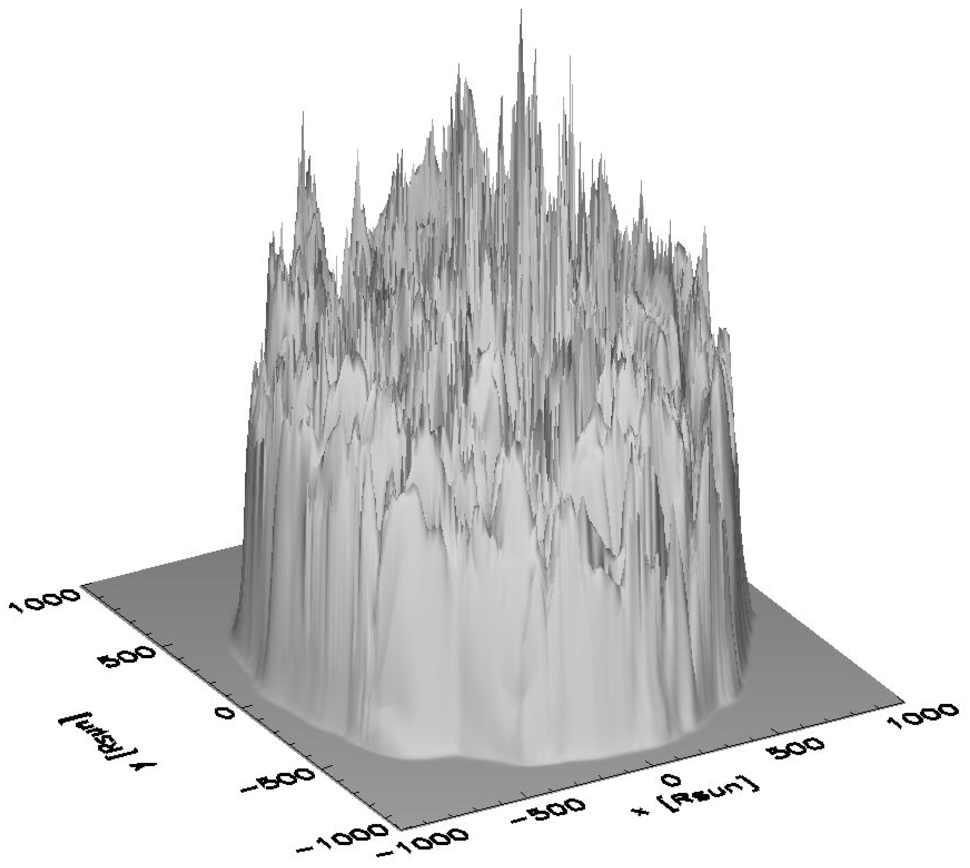}
      \includegraphics[width=0.5\hsize]{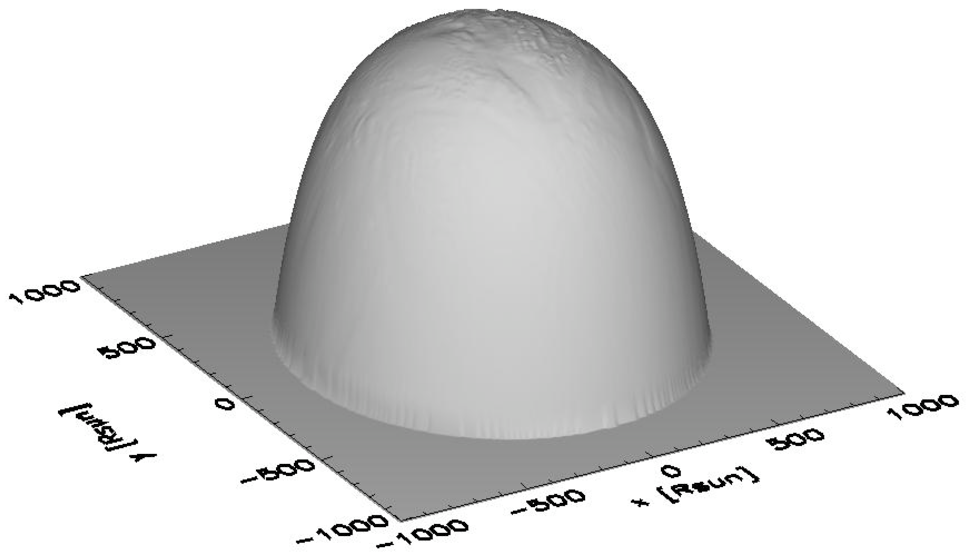}\\
 \includegraphics[width=0.5\hsize]{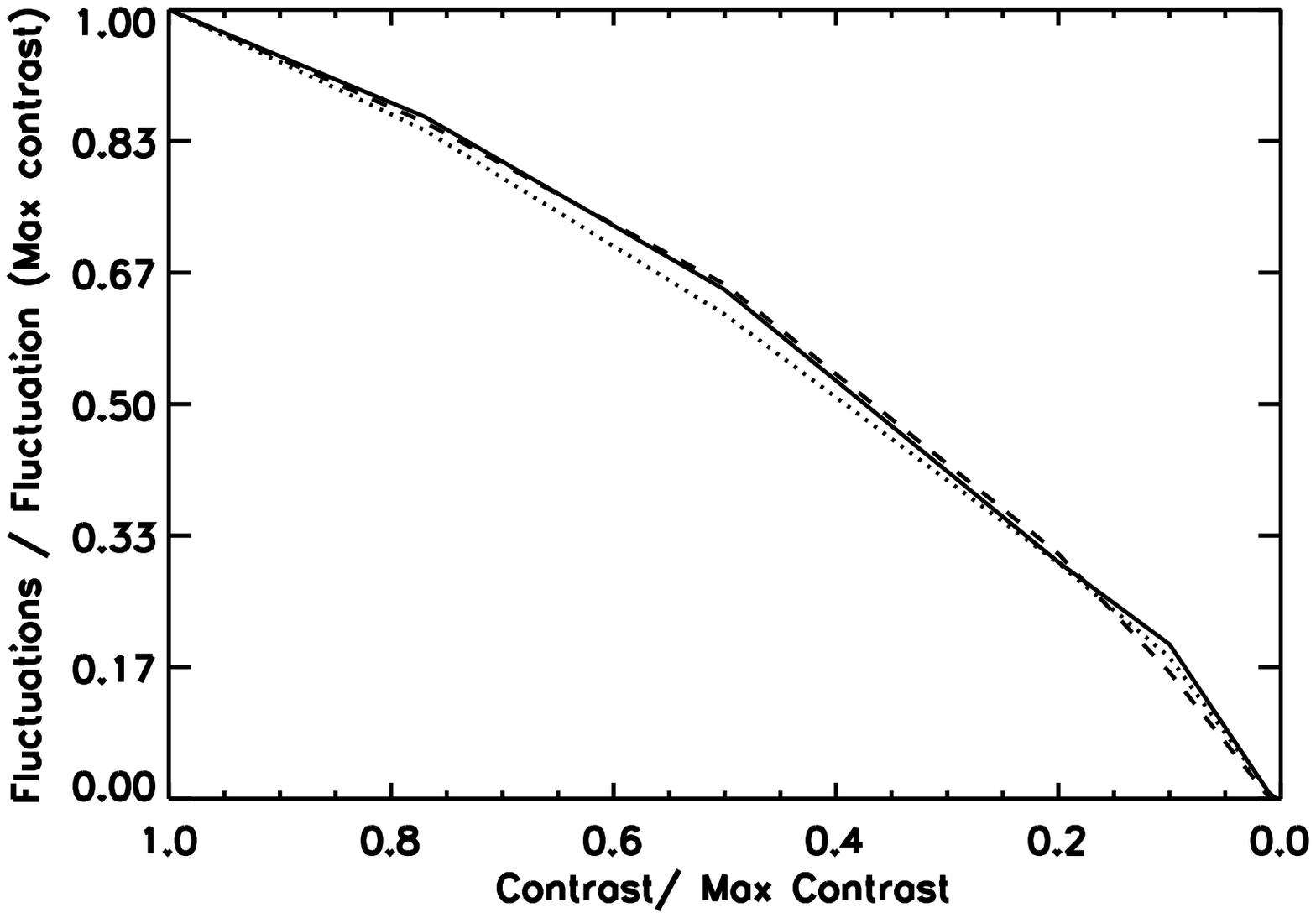} 
    \includegraphics[width=0.5\hsize]{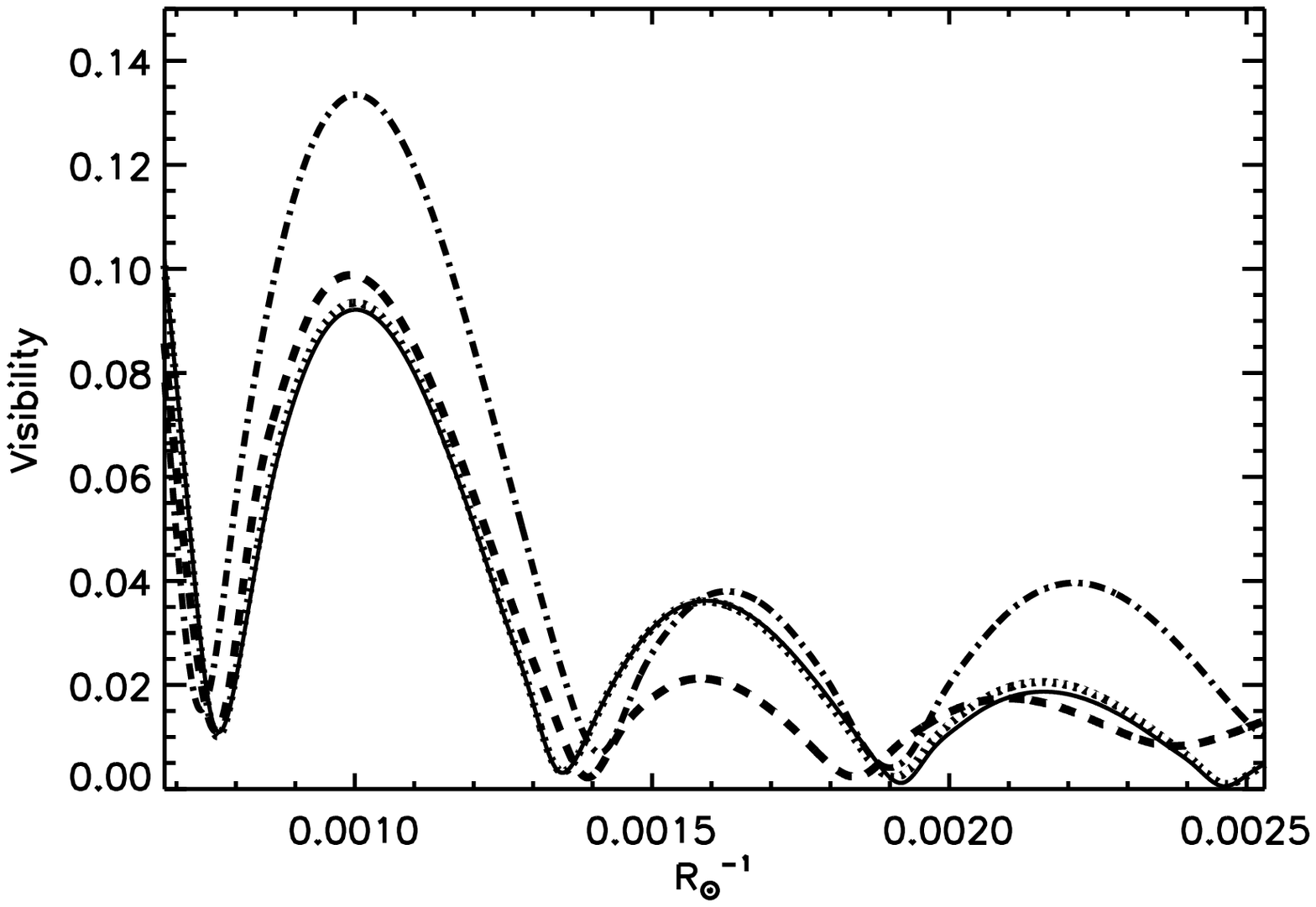}
      \end{tabular}
      \caption{ \emph{Top left panel: }three-dimensional image of a snapshot with nominal intensities. 
      \emph{Top right panel: }same snapshot with a feature contrast reduced to 1$\%$. 
      \emph{Bottom left panel: } standard deviation of the visibility curves at 36 angles 5$^\circ$)
       apart, for decreasing feature contrast. 
       Solid line is for the top of the second lobe ($\sim 0.0010~{\rm R}^{-1}_\odot$), dashed line is 
       for the top of the third lobe ($\sim 0.0016~{\rm R}^{-1}_\odot$), and dotted line is for the top of 
       the fourth lobe ($\sim 0.0022~{\rm R}^{-1}_\odot$). 
       \emph{Bottom right panel: }Visibility in the second, third, and fourth lobes
        for one particular position angle. 
       The dot-dashed line shows the original simulation contrast.  The dashed line, and the dotted line
       show the visibility with a feature contrast reduced to 50$\%$, and  1$\%$ respectively.
       The solid line is the fitted LD profile computed for this snapshot.}
         \label{surface_contrast}
   \end{figure*}

It may however turn out that approximations in our modelling (e.g., limited spatial resolution, 
grey radiative transfer) significantly affect the intensity contrast of the granulation.
Indeed, the radiation transfer in our RHD models uses a frequency-independent grey 
treatment to speed 
up the calculations. This approximation leads to errors in the mean temperature structure
 in the optically thin layers that are difficult to quantify. The implementation 
of non-grey opacities can decrease the temperature fluctuations compared to the grey case 
\citep[e.g.,][for local RHD models]{1994A&A...284..105L}. As a consequence, 
the intensity contrast will be decreased, reducing the visibility fluctuations. 
To investigate its impact  on visibilities, we artificially decrease  the intensity contrast on one of our images. 
we use the snapshot with its  nominal intensities as reference. We first fit a LD law (as in Sect~\ref{LDsect})
to the radially average intensity profile. After subtracting this average profile from the intensity map
we are left with the fluctuations caused by granulation.
 We measure the contrast 
$C_{\rm{ref}}$=$\frac{I_{\rm{max}}-I_{\rm{min}}}{I_{\rm{max}}+I_{\rm{min}}}$. It is then easy to scale that 
contrast before adding again the LD profile, to recover a reduced contrast image.
 An example of the resulting intensities is shown in Fig.~\ref{surface_contrast} (top row). 
At a  contrast of only 1$\%$ of the nominal one, small surface structures 
are hardly visible. 
As previously, we determine  $\sigma_{\rm{vis}}/\rm{vis}$ for all the images with various contrats,
around the top of  the second 
($\sim$0.0010~R$^{-1}_\odot$), third ($\sim$0.0016~R$^{-1}_\odot$), and fourth lobes
($\sim$0.0022~R$^{-1}_\odot$). The bottom left panel of Fig.~\ref{surface_contrast} shows that when
the contrast is reduced, and the surface structures fade out, the resulting visibility fluctuations 
decrease similarly in  all the lobes (almost proportionally to the intensity contrast decrease). 
Reducing the contrast brings of course the visibility curves towards the visibility of the LD profile 
(Fig.~\ref{surface_contrast}, bottom right panel). This proportionality 
can be used to determine the granulation contrast from observations of  the visibility 
fluctuations with time or position angle. 

\subsection{Importance of spectral resolution in interferometry: the H and K bands}\index{Differential inteferometry}\label{diff_vis}

\begin{figure*}
  \centering	
      \begin{tabular}{c}
      	\includegraphics[width=1.0\hsize]{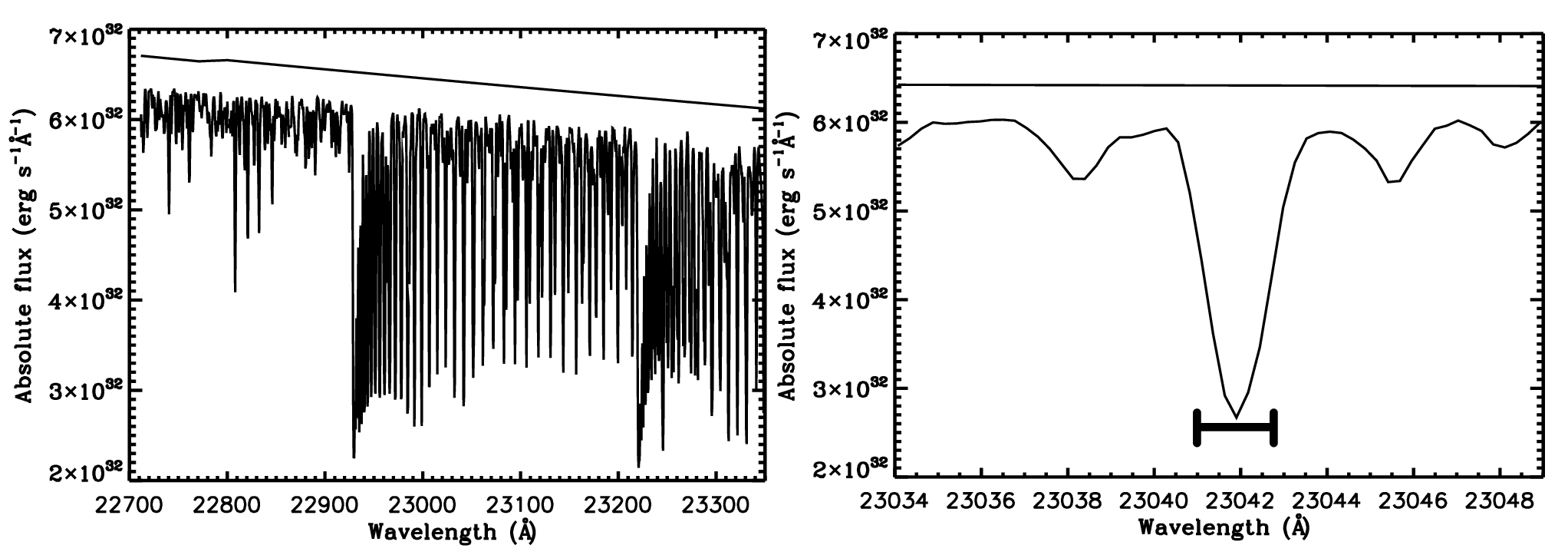} \\
  	\includegraphics[width=1.0\hsize]{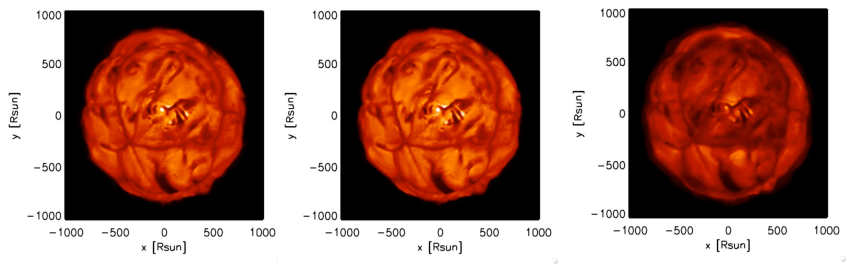} \\
	\includegraphics[width=0.7\hsize]{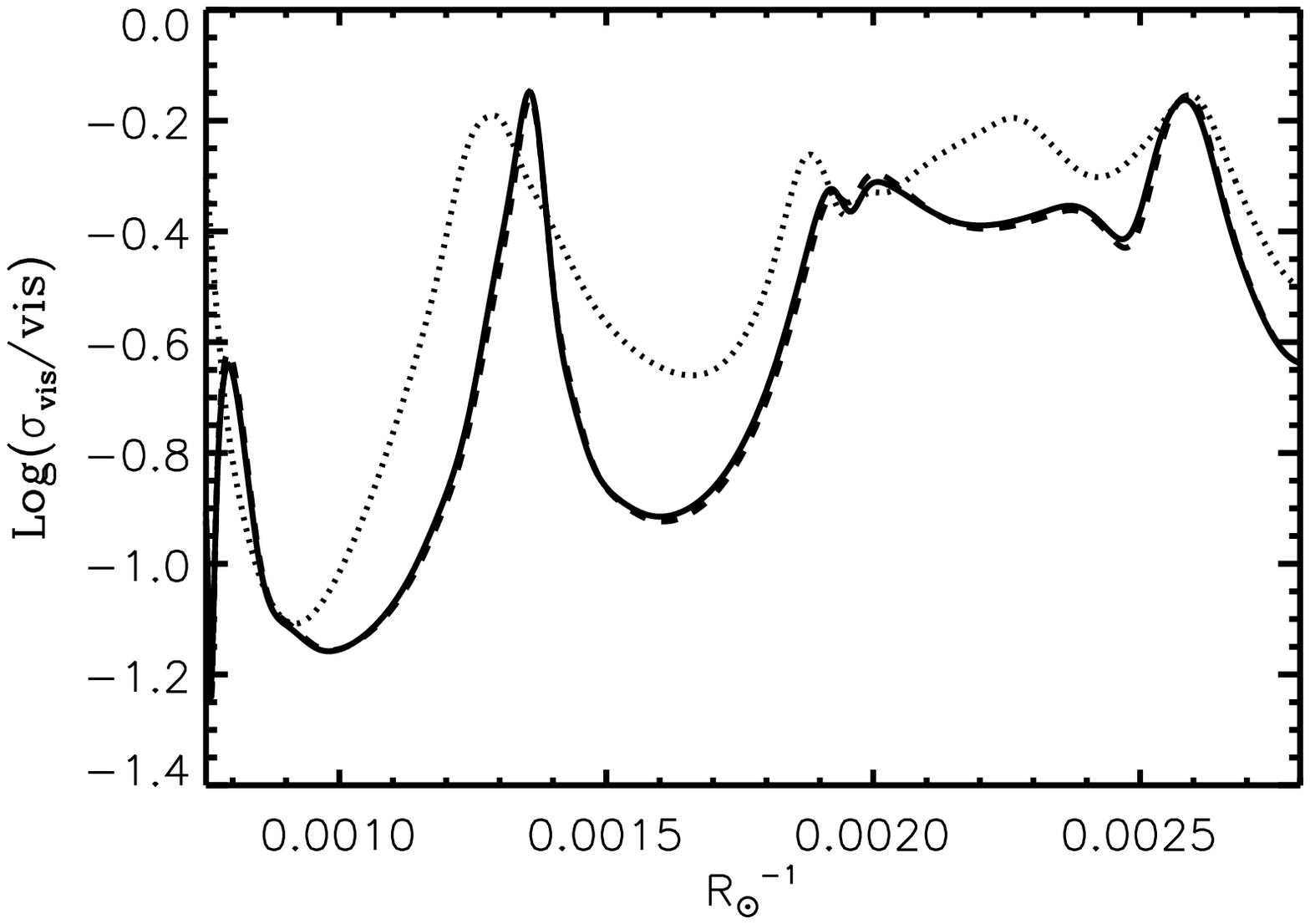} 
	 \end{tabular}
      \caption{\emph{Top row:} synthetic spectrum centered on the CO first overtone band head. 
      The left panel shows the range of wavelengths spanned by one resolution element at the
       VLTI-AMBER low spectral resolution of 35. The right panel shows the same  for the 
       VLTI-AMBER medium spectral resolution of 1\,500,  and the high spectral resolution of 12\,000 (thick mark). 
       \emph{Central row:} intensity maps  for those three spectral resolution elements. 
       The intensity range is [0;$10^5$]\,erg\,cm$^{-2}$\,s$^{-1}$\,{\AA}$^{-1}$. 
       \emph{Bottom row:} standard deviation of the visibility in the  second, third and fourth lobes
        Solid and dashed lines correspond low and medium resolution respectively, and the dotted line to 
        high resolution.
         }
         \label{amber_resolution}
  \end{figure*}

Interferometric observations done through a broad band filter blend information from the lines and continuum.
Spectral resolution allows to recover much richer information, both from visibility moduli and phases. 
The VLTI-AMBER interferometer provides spectral resolutions of R=35, 1\,500, and 12\,000. 
In order to show the differences between these resolutions, we compute intensity maps around the CO 
first overtone line at 23041.75~\AA \ (log($gf$)=-5.527 and $\chi_{ex}$=0.180~eV) for the three resolutions 
(Fig.~\ref{amber_resolution}). The resulting images are shown in the central row of the Figure, and the spectrum
in the top row. The contrast, defined as in previous Section, is similar for the low and medium resolution 
images but it is $\sim30\%$ lower in the high resolution image, at the CO line wavelength. 
Large fluctuations are seen in all lobes, but they are smaller than those seen in the H band IONIC 
filter (see Fig~\ref{fluct_vis_2}). The visibility fluctuations for the high spectral resolution image 
in the CO line are larger (second, third and fourth lobes; see dotted line in bottom panel), 
despite a lower intensity contrast, presumably because of the darkening of large patches of
the simulated stellar surface.

We also computed wavelength dependent visibility curves in the H band for the high and medium VLTI-AMBER
resolutions. Fig.~\ref{diff_int} displays a three-dimensional view of the visibility curves with a 
resolution of 12\,000, and 1\,500 (top panels). The simulated star has been scaled to an apparent diameter 
of $\sim$43.6~mas ({the observed diameter of $\alpha$~Ori \citealp{2004A&A...418..675P}). 
The displacement of the zero points with wavelength is easily seen, as well as the amplitude variations
in the higher frequency lobes. 
 
In order to mimic differential observations with an interferometer at medium and high spectral 
resolution, we also show in Fig.~\ref{diff_int}, the variation of the visibility modulus 
with wavelength for a fixed baseline (15m, i.e. in the second lobe, at $\nu=45{\rm arcsec}^{-1}$).
The visibility shows variations correlated with the flux spectrum: it decreases in absorption lines. 
In fact, at these wavelengths wiggles and dark spots appear on intensity maps (Fig.~\ref{amber_resolution},
central right panel) increasing the visibility signal at frequencies higher than the second lobe. 
The visibility variations are much attenuated at lower spectral resolution.   
Observations at wavelengths in a spectral line,
and in the nearby continuum will probe different atmospheric depths, and thus layers at
different temperatures. They will thus provide important information on the wavelength dependence of 
limb darkening. Moreover, 
as the horizontal temperature and density fluctuations depend on the depth in the atmosphere, differential 
observations, with relative phase determination will provide unique constraints on the granulation pattern.
The visibility variations in Fig~\ref{diff_int}, 
such as the  steep visibility jump from 0.123 to 0.107 between 1.5975 and 1.5980 $\mu$m,
could be measured in differential interferometric mode at high spectral resolution with the current precision
at VLTI-AMBER (1$\%$ for visibilities of $\sim$5-10$\%$), with optimal sky conditions. 

\begin{figure*}
   \centering
    \begin{tabular}{cc}
       \includegraphics[width=0.5\hsize]{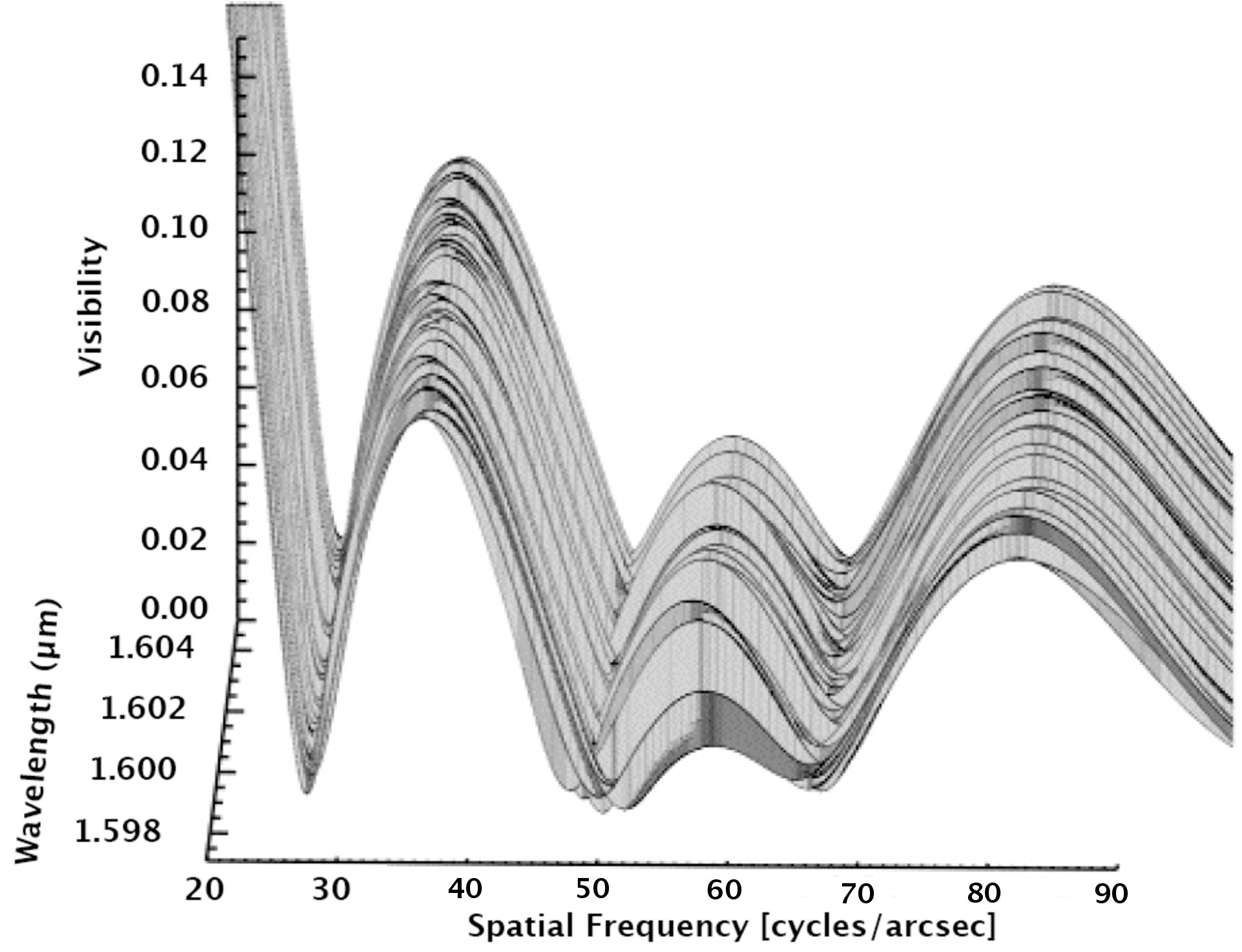} 
        \includegraphics[width=0.5\hsize]{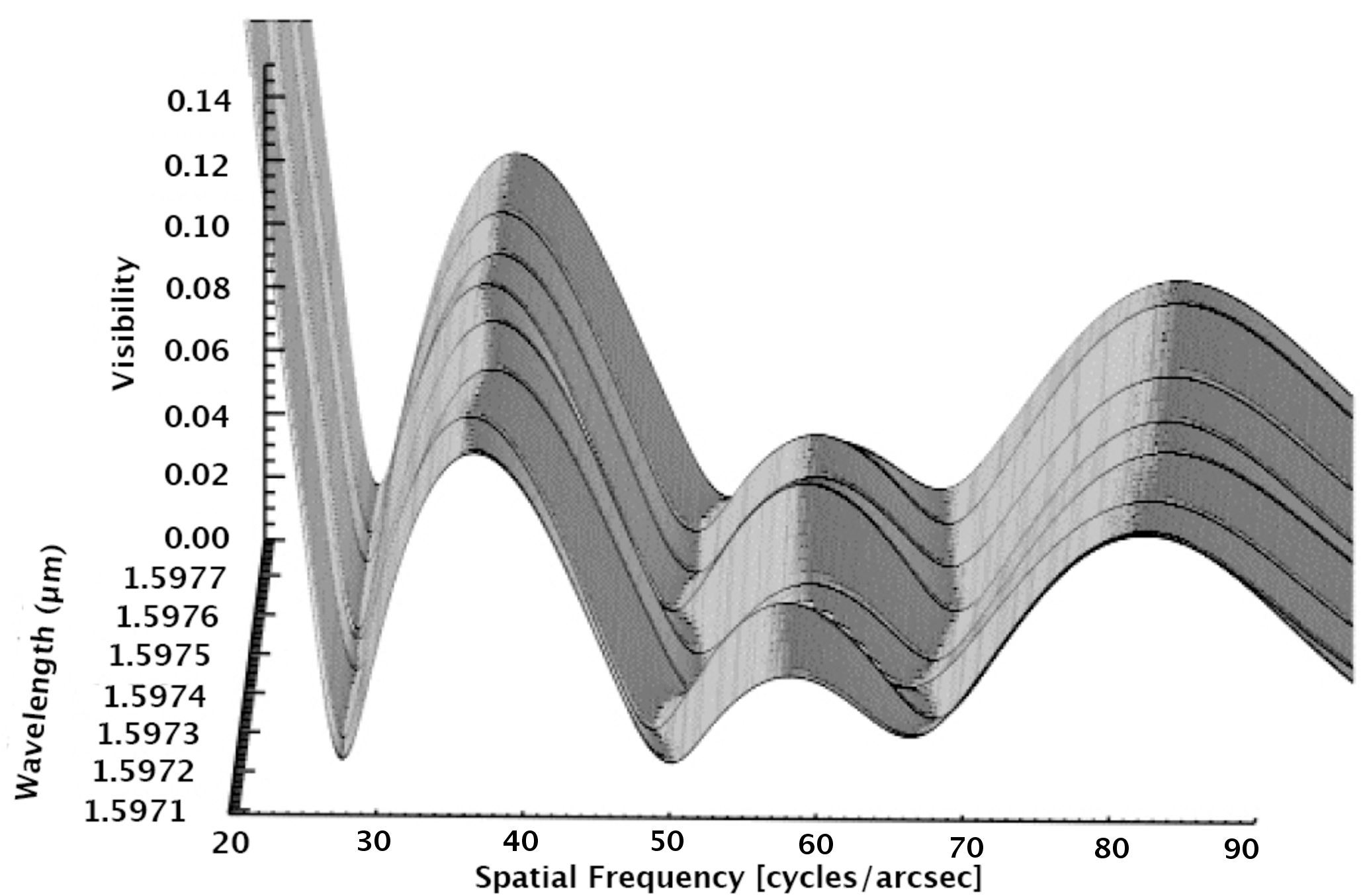} \\
 \includegraphics[width=0.55\hsize]{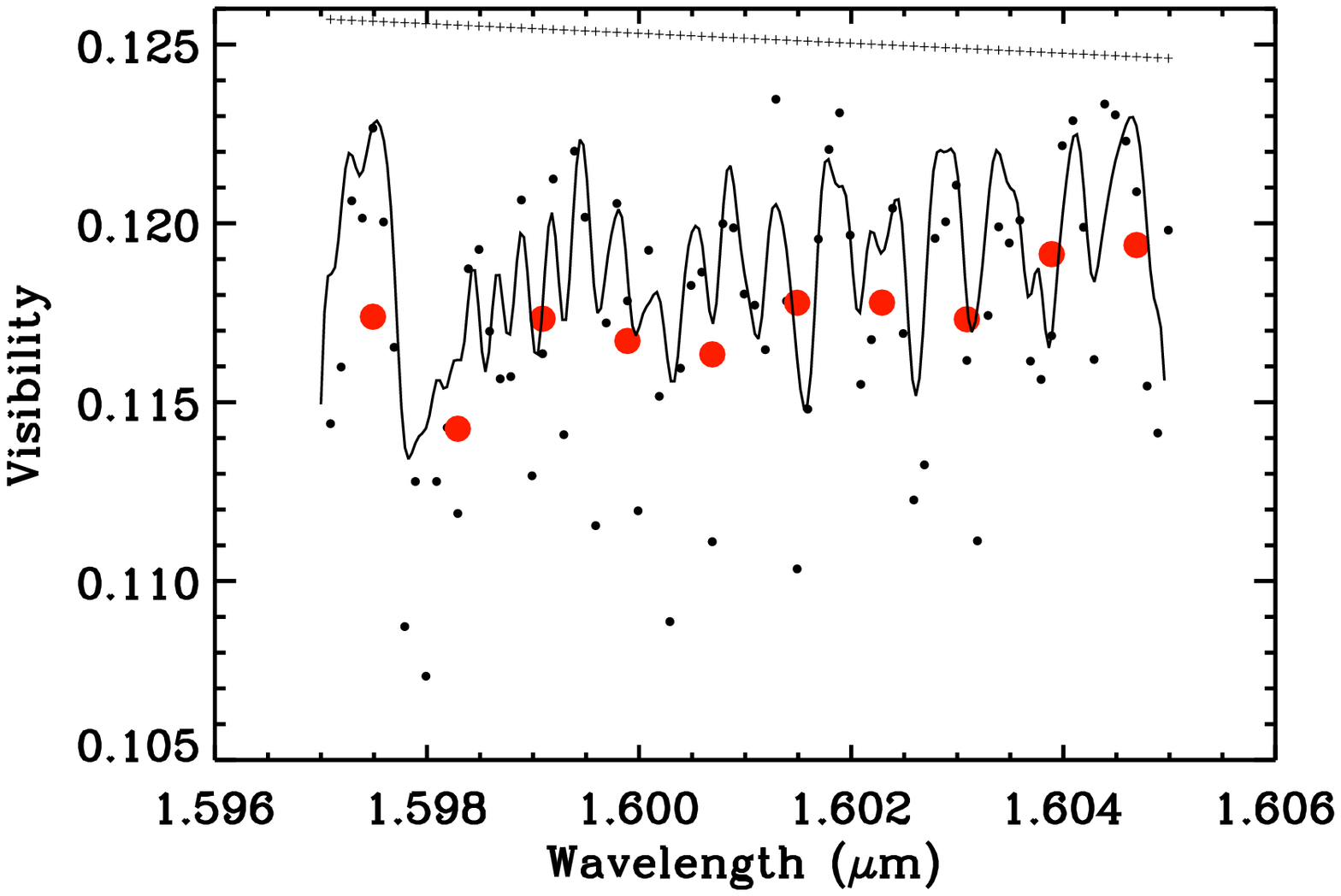} 
  \end{tabular}
      \caption{\emph{Top left panel:} three-dimensional view of the visibility curves as a function of wavelength
     for a particular position angle.
       The spectral resolution is 12\,000. \emph{Top right panel:} same as in top left panel at a spectral 
       resolution of 1\,500. \emph{Bottom panel:} Visibility as a function of wavelength 
      for a baseline of $\sim$15 m, i.e. the top of the second lobe, at one particular position angle.
      The simulation has been scaled to an apparent diameter of $\sim$43.6 mas.
       The synthetic spectrum convolved to a resolution of 12\,000 is over-plotted (thin solid line). 
      The small black dots correspond to the highest resolution with AMBER (12\,000), while the big red dots 
      correspond to the medium resolution (1\,500). The crosses show the uniform disk of 43.6 mas. 
       When changing the position angle, the expected standard deviation is about 10$\%$ of the visibility 
       (see Fig.~\ref{fluct_vis_2}).
       }
         \label{diff_int}
   \end{figure*}

\subsection{Closure phase: departure from circular symmetry}\label{clph}

As terrestrial atmospheric turbulence affects the phases of the  complex visibilities with 
random errors, it is impossible to derive them for individual pairs of telescopes. 
Instead one uses closure phase between three telescopes, as the sum of all phase differences
removes the atmospheric contribution, leaving the phase information of
the object visibility untouched \citep[see e.g.,][]{2007NewAR..51..604M}.
The closure phase is thus an important complementary piece of information, which can reveal 
asymmetries of RSG atmospheres.
Fig.~\ref{clphaFig} shows the scatter plot of the closure phase 
of one snapshot of the RHD model computed in the IONIC filter (the scatter is similar for the 
K222 filter). The behavior is similar for all the snapshots. We used 500 random baseline triangles 
with a maximum linear extension of 40 m, and plot the closure phase  as a function of the
triangle  maximum baseline. The closure phases deviate from zero or $\pm\pi$ already at $\sim$10 m 
(0.0008~{\rm R}$^{-1}_\odot$ if we scale the model to an apparent diameter of 43.6~mas at a distance of 174.3~pc).
At higher baselines it is clearly different from zero or $\pm\pi$, 
values which indicate a point symmetric brightness distribution.
This is a clear signature of surface inhomogenities. The characteristic size distribution on the stellar 
surface can also be derived from the closure phase:  the contribution of small scale convection-related 
surface structures increases with frequency. The first deviation at $\sim$0.0008~{\rm R}$^{-1}_\odot$ 
(just beyond the first zero, see Fig.~\ref{vis_zoom2}) corresponds to the deviation from circular 
symmetry of the stellar disk.
It may be very efficient to constrain the level of asymmetry of RSG atmospheres
by accumulating statistics on closure phase at short and long baselines, as they are easily measured with 
great precision. Small departure from zero will immediately reveal departure from symmetry.

\begin{figure}
   \centering
 \includegraphics[width=1.0\hsize]{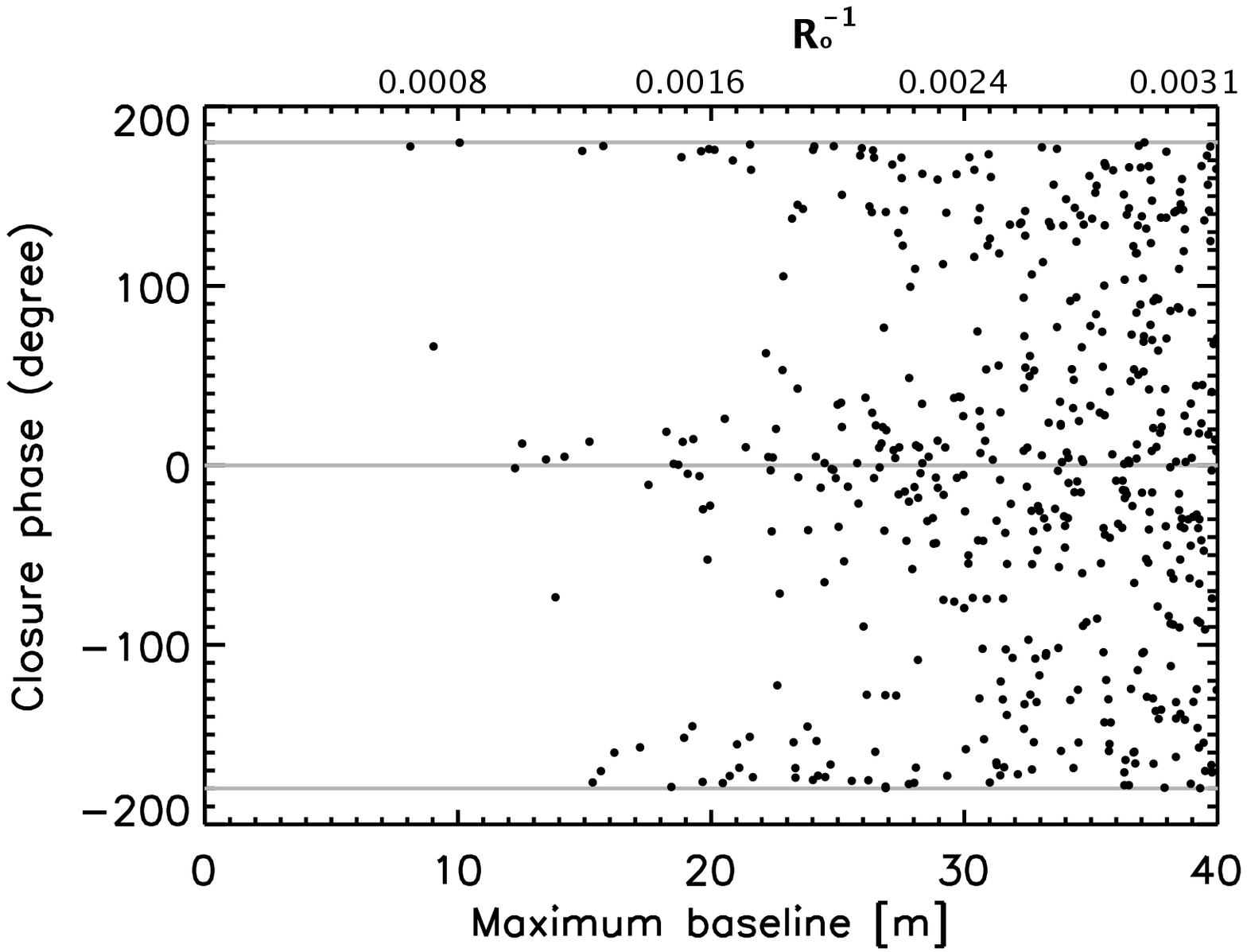}  
      \caption{Scatter plot of closure phases in the IONIC filter centered at 1.64~$\mu$m of 500 random 
      baseline triangles with a maximum linear extension of 40 m.
       Closure phases are plotted
      against  the longest baseline of the triangle. 
      The upper x-axis corresponds to synthetic observations of the simulation at an apparent 
      diameter of 43.6~mas (which corresponds to $\alpha$\ Ori at a distance of 174.3\,pc). 
      The axisymmetric case is represented by the grey lines.}
         \label{clphaFig}
   \end{figure}

We also computed the closure phase for the different K band VLTI-AMBER spectral resolution intensity 
maps of Fig.~\ref{amber_resolution}. The large deviations from circular symmetry are already noticeable 
at low spectral resolution (Fig.~\ref{clphares}, left panels) and the closure phase scatter do not differ
 much from the high spectral resolution one (right panel). This offers prospects of detecting asymmetries due to granulation
without resorting to high spectral resolution.

\begin{figure*}
   \centering
    \begin{tabular}{cc}
 \includegraphics[width=0.5\hsize]{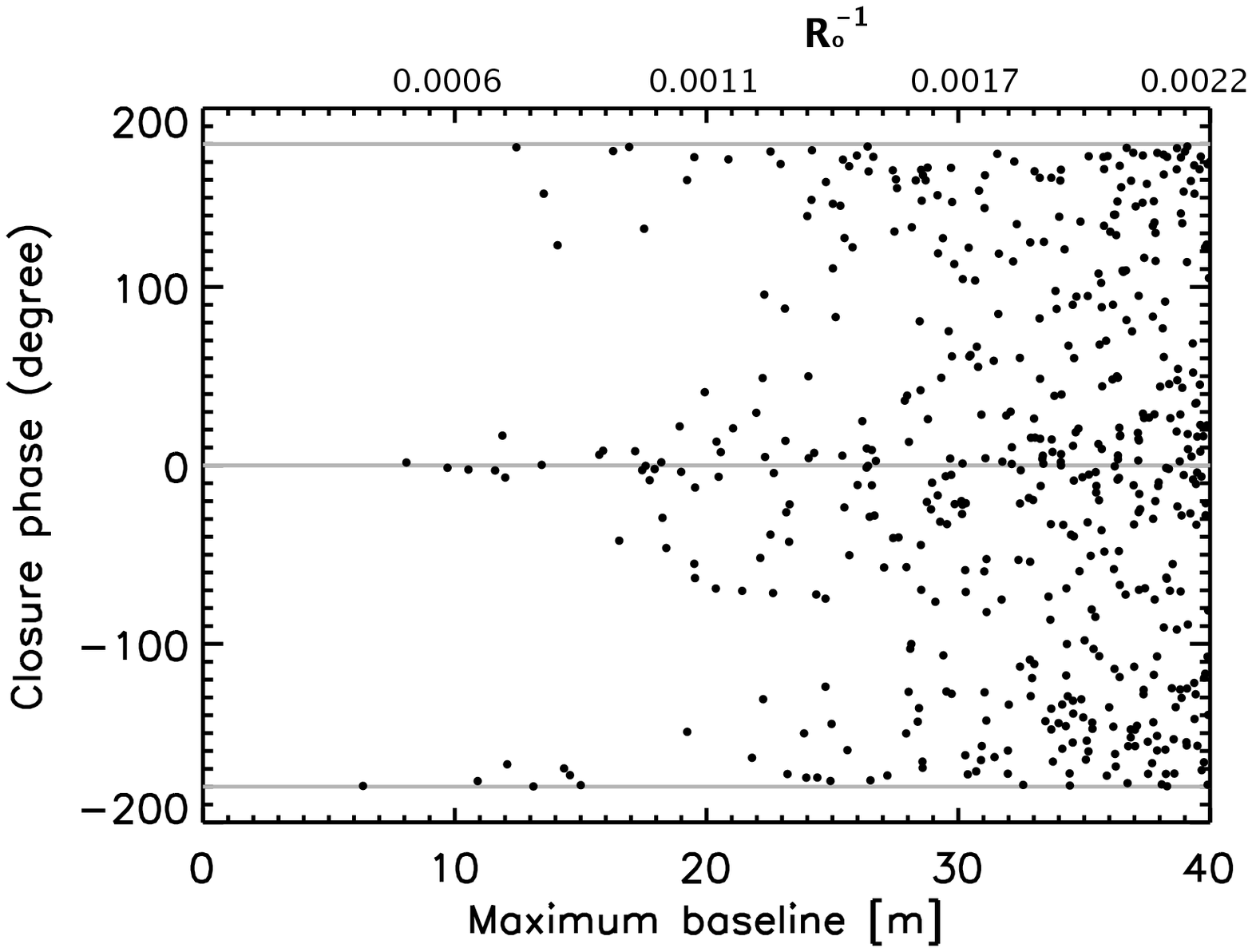}  
\includegraphics[width=0.5\hsize]{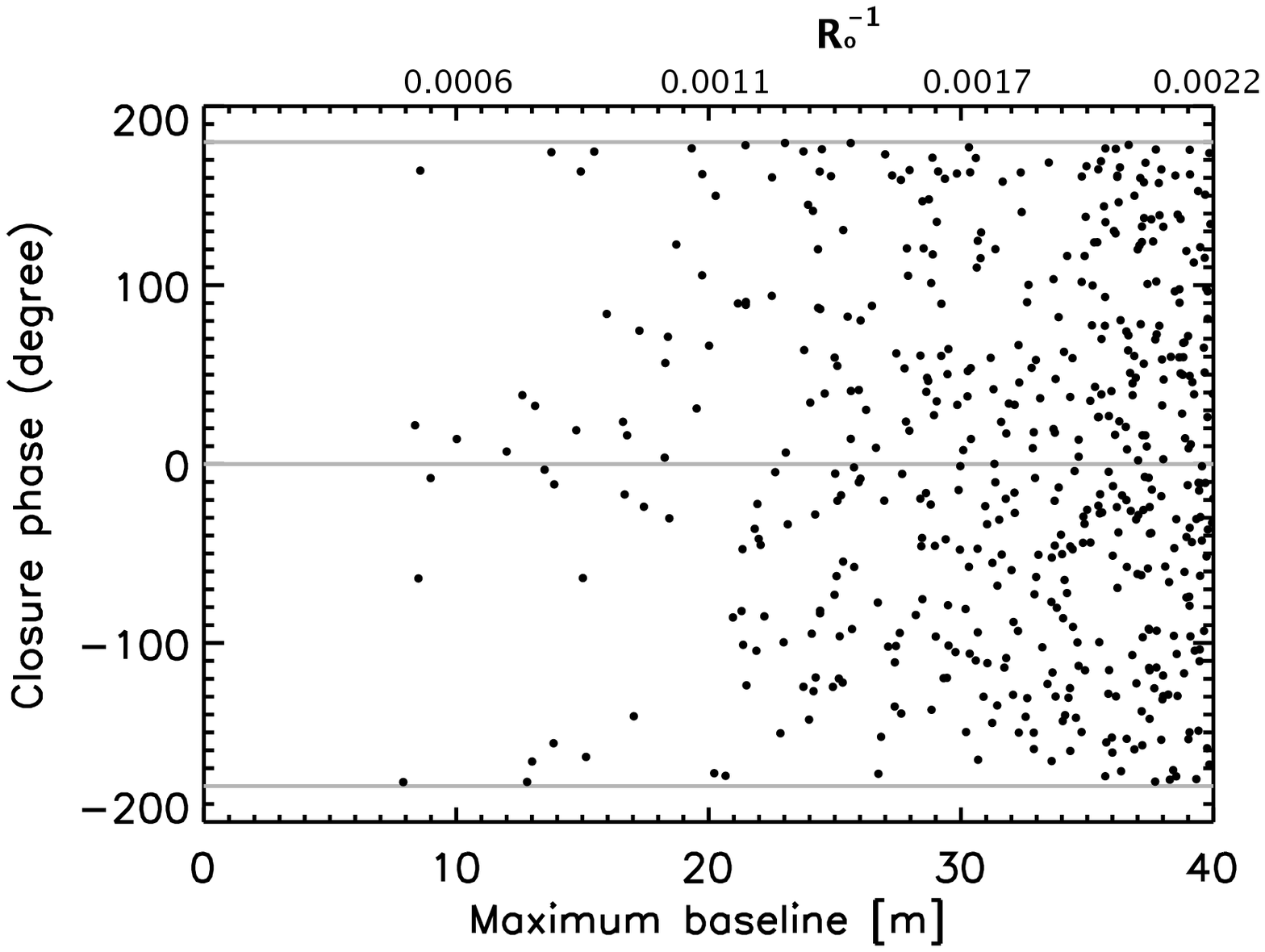} 
  \end{tabular}
      \caption{Scatter plot of closure phases (cf~Fig.~\ref{clphaFig} for details) obtained from
       the VLTI-AMBER K band low, and high spectral resolution intensity maps (Fig.~\ref{amber_resolution}, 
      central left, and right panels).}
         \label{clphares}
   \end{figure*}


\section{Comparison with interferometric observations of $\alpha$~Ori}\label{aOri-result}

A first confrontation of our model predictions to real observations is possible for $\alpha$~Ori.  
We compare the synthetic visibilities derived from our RHD simulations in the continuum filter K222 
(Fig.~\ref{filterFLUOR}) with the observation of $\alpha$~Ori by \cite{2004A&A...418..675P} 
that reach the third lobe in the K band.
The absolute model dimensions have been scaled to match the interferometric observation in the 
first lobe. This corresponds to an apparent diameter of 43.6~mas at a distance of 179~pc. 
These values are in agreement with \citeauthor{2004A&A...418..675P}, who found a diameter 
of $43.64\pm0.10$~mas, and \citet{2008AJ....135.1430H}, who reported a distance of $197\pm45$~pc. 

We computed over 2\,000 visibility curves, and we find that the data are 
within the visibility fluctuations due to the granulation of the simulation 
(Fig.~\ref{comparison_aori}), as already shown in \citealp{2007sf2a.conf..447C}. Within this 
large number of visibility curves, we find  some that match all the observation points better 
than the uniform disk (with a diameter of 43.33 mas; \citealp{2004A&A...418..675P}), or 
limb-darkened disk model (linear limb darkening law, $I\left(\mu\right)=1-a\left(1-\mu\right)$, 
with a diameter of 43.64 mas and $a=0.09$ also in \citeauthor{2004A&A...418..675P}). 
See Fig.~\ref{comparison_aori}. The best match has a reduced $\chi^2$=0.21,
 and all the visibility curves fall within a $\chi^2$ range of [0.21,18.1]. 
 Our RHD simulations are a great improvement over parametric models 
 (the UD model with reduced $\chi^2$=19.9, and the LD model with $\chi^2$=22.3) 
 for the interpretation of these interferometric observations. 
 The observations points in the first, second, and third lobes can be reproduced with a single
 visibility curve, from the projection at a particular position angle of one of our snapshots (see Fig.~\ref{comparison_aori}). 
 There is one observed point in the first lobe at 24.5 arcsec$^{-1}$ which is difficult 
 to reproduce. In fact, adjustments on the absolute model dimensions of the star in order 
 to fit this point, 
 would lead to mismatch of the other observations at higher frequencies. 
 However, this may be a problem with the calibration of the observation. 
 
A more detailed comparison with $\alpha$~Ori data in the H band \citep{2006sf2a.conf..471H} will 
be presented in a forthcoming paper (Chiavassa et al. 2009, in prep.)

\begin{figure*}
   \centering
    \begin{tabular}{cc}
    \includegraphics[width=0.5\hsize]{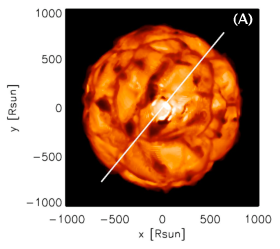} \\
 \includegraphics[width=0.5\hsize]{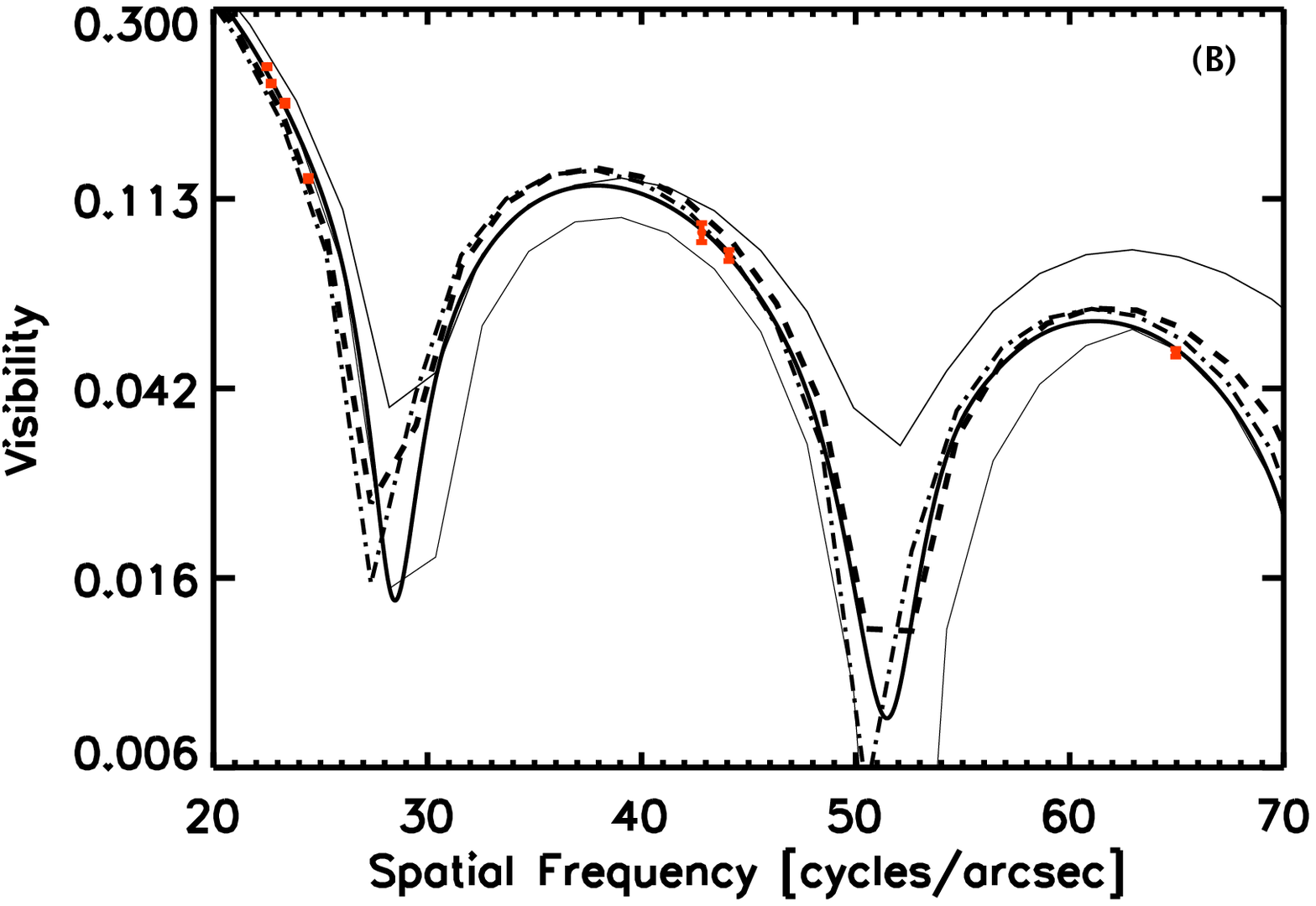} 
    \includegraphics[width=0.5\hsize]{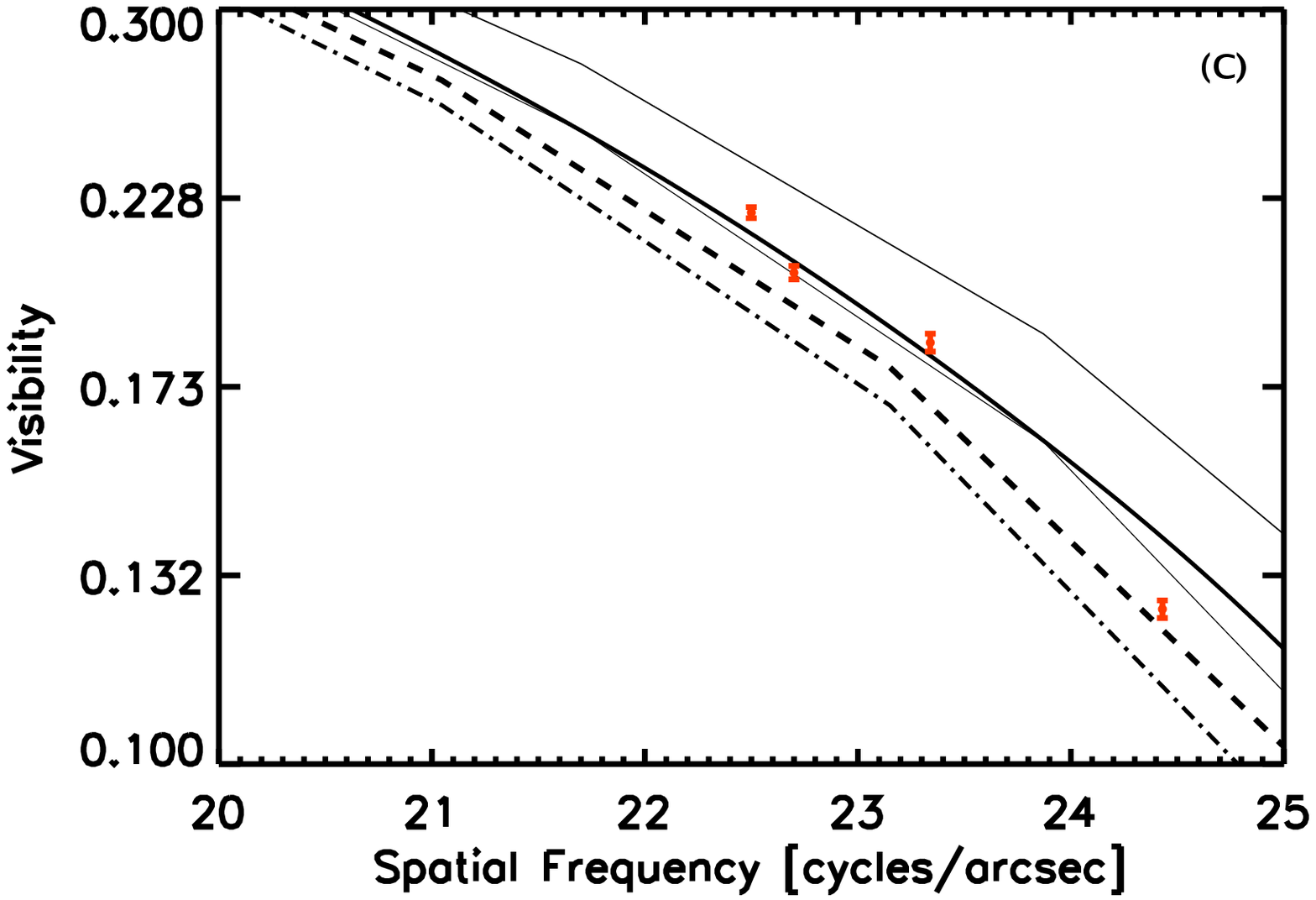}\\
    	\includegraphics[width=0.5\hsize]{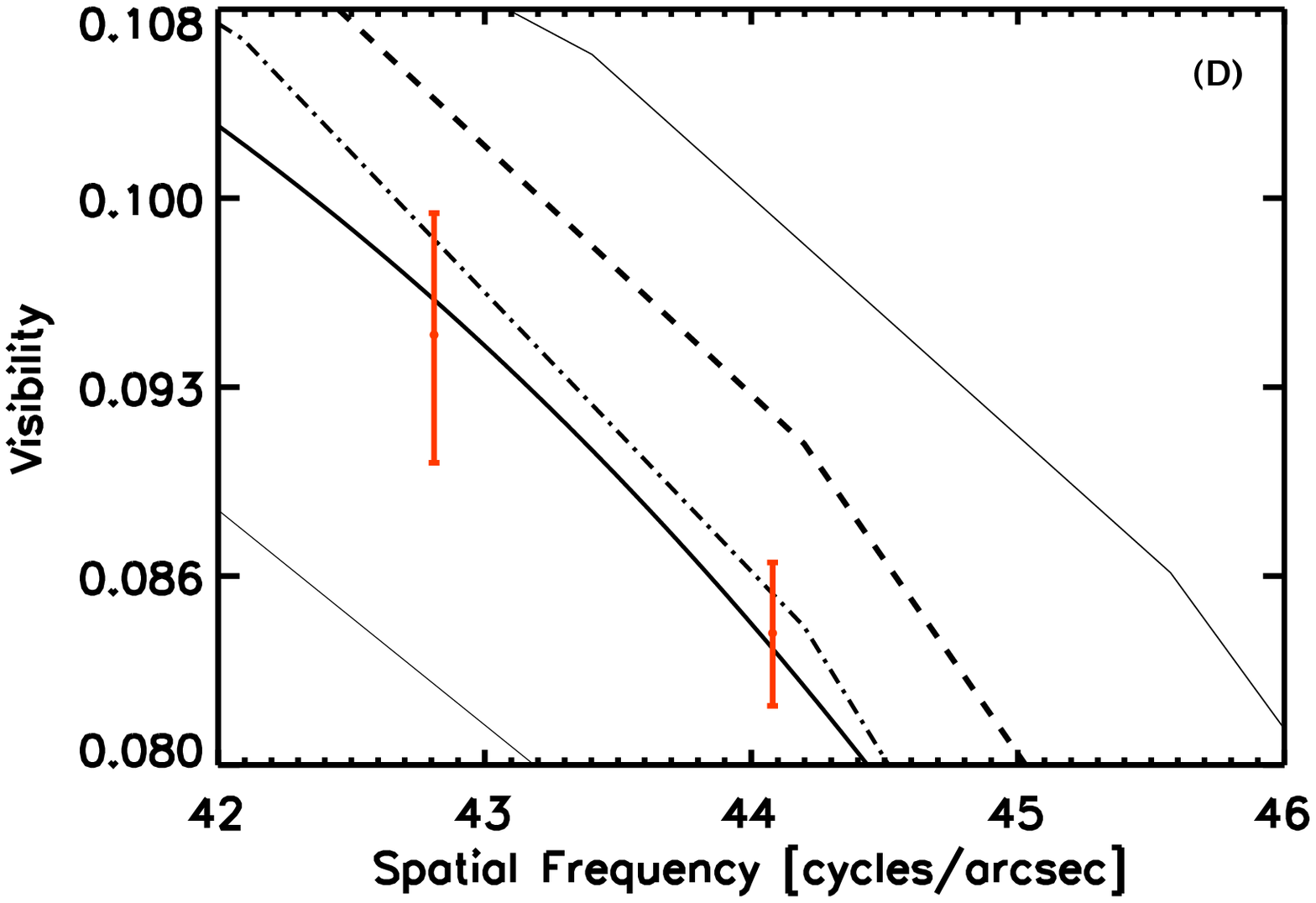} 
    \includegraphics[width=0.5\hsize]{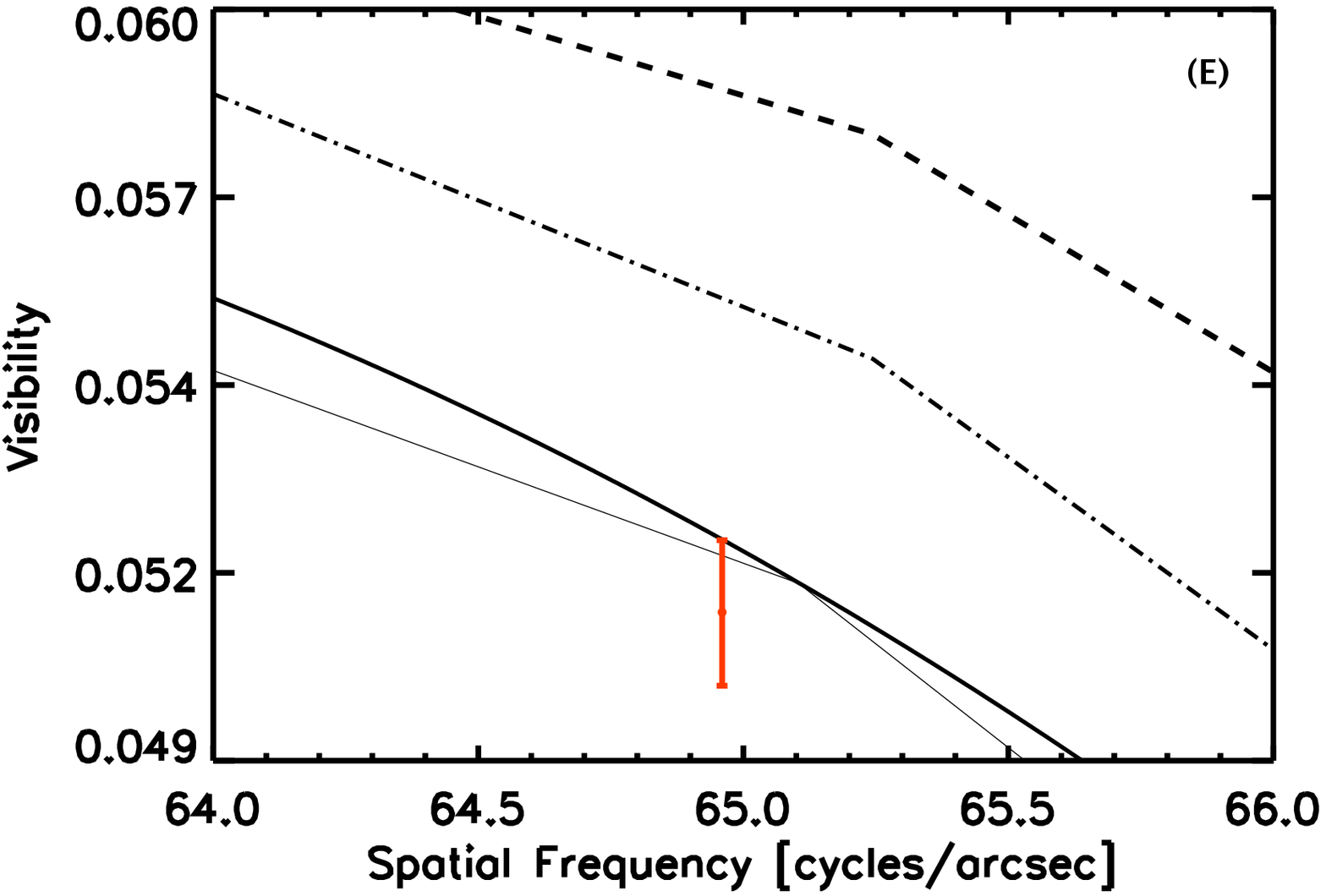}\\
    \end{tabular}
     \caption{Comparison of the RHD simulation with the $\alpha$~Ori observations (red dots with error bars) 
     by \cite{2004A&A...418..675P}. \emph{Top panel:} intensity map in the K222-FLUOR filter of the 
     snapshot that best matches the interferometric data. 
     The range is [0;$10^5$]\,erg\,cm$^{-2}$\,s$^{-1}$\,{\AA}$^{-1}$. 
     The stellar parameters of this snapshot are: $L = 93\,480\,{\rm L}_\odot$, $R = 833\,{\rm R}_\odot$,
     $T_{\rm eff} = 3497$\,K and $\log~g = -0.34$. The simulation has been scaled to an apparent diameter 
     of 43.6~mas at a distance of 179~pc. The white line indicates the position angle of the 
     projected baselines.  
     \emph{Other panels:} Synthetic visibilities from the simulation, compared with
     observations. Panel B) covers the whole observational range, and panels
     C)-E) are close-ups of the clusters of observations. The thick solid line corresponds to the best 
     match visibility curve(reduced $\chi^2$=0.21). The thin solid lines show the minimum and maximum 
     extent of variations of the visibilities. The dot-dashed, and the dashed lines are
     the LD, and the UD models
     used by \citeauthor{2004A&A...418..675P} (reduced $\chi^2$=22.3, and 19.9). 
     Note the logarithmic visibility scale.}
              \label{comparison_aori}
\end{figure*}

\section{Conclusions}

Our radiation hydrodynamics simulations confirm that only a few large
     granules cover the surface of RSG stars. The granules of the simulation
     we analyze here are 400--500\,R$_\odot$ in diameter, and have
     lifetimes of years. Smaller scale structures develop and evolve within these
     large granules, on shorter timescales (a month).
     
  We demonstrate that  RHD simulations are necessary for a proper quantitative analysis 
   of interferometric observations
 of the surface of RSGs beyond the smooth, symmetrical, limb-darkened intensity profiles. 
 We give new average limb darkening coefficients within the H and K bands, 
 that are significantly different from commonly used UD or LD profiles.
 However, these LD coefficients fluctuate with time, and the average is only indicative.
Our model surface granulation causes angular and temporal variations of 
visibility amplitudes and phases.
 In the first lobe, sensitive to the radius, fluctuations can be as high as  5$\%$, and radii
 determinations can be affected to that extent: the radius determined with a UD fit 
 is 3-5$\%$ smaller than the radius of  the simulation, while the radius determined with a 
 fully LD model is $1\%$ smaller.
 
The second, third, and fourth lobes, that carry the signature of limb-darkening, 
and of smaller scale structure, are very different from the simple LD case. The visibility amplitudes
can be greater than the UD or LD case, and closure phases largely differ from 0 and $\pm\pi$, due to 
the departure from circular symmetry.
The visibilities also show  fluctuations with position angle, and with time, that are directly 
related to the granulation contrast.
 We also want to stress that high spectral resolution provides an extremely valuable information.
 The stellar surface appears dramatically different in an absorption line and in the nearby
 continuum, and differential observations should be easier to carry
 out with high precision.
 At lower resolution (e.g. $R=1\,500$) continuum and line information get mixed and there 
 is a considerable loss of differential signal.
    
We conclude that the detection of the signature of granulation, as predicted by our simulation, is 
measurable with todays interferometers, if observations of both amplitude and closure phase
are made at  high spatial frequencies (second, third, and fourth lobes, or even further if possible). 
These observations will 
provide direct information on the time scale of variation, and of the size and contrast of granulation. 
 
 A few RSGs are prime targets for interferometry, 
 thanks to their large diameter, proximity, and to a high infrared luminosity. Only 4 or 5 
 are sufficiently close and bright that imaging can be attempted, but a larger number (10-20) are
 within reach of less ambitious programs, like closure phase measurements.

 Three approaches can be combined to characterize the granulation pattern:
\begin{itemize}
\item searching for angular visibility variations, observing with the same telescope configuration 
(covering high spatial frequencies) and using the Earth rotation in order to span 6-7 position angles 
in one night should be enough, if measurement errors can be kept below 10$\%$, for visibilities of the
 order of 5 to 10$\%$. One or two other telescope configurations would provide more frequency points,
 but then the change of configuration must be made within days, which is actually possible at VLTI;
\item looking for temporal visibility fluctuations by observing at two (preferably more) epochs 
$\sim$1 month apart with the same telescope configuration. This can be easily scheduled on existing
interferometer like CHARA, or the VLTI;
\item looking for visibilities as a function of wavelength, at high spectral resolution, in 
different spectral regions belonging to spectral features and continuum. If measurement errors 
can be kept close to 1$\%$, for visibility of the order of $\sim$10$\%$, variations of the 
visibility correlated with the flux spectrum could be detected, indicating variations of the radius, 
of the limb-darkening, or of the granulation pattern. Such relative  measurements are more easily done
at the required precision  than absolute visibility measurements. 
\end{itemize}

These observations will bring us a wealth of information on the stars, but also on our RHD models.
We know they suffer from limitations. The confrontation to observation will help us
decide what approximations must be relaxed.
The simulations are primarily constrained by execution time, which
     neccesitates several approximations, the most important ones being the
     limited spatial resolution and the complete lack of wavelength resolution,
     i.e., grey radiative transfer. This speeds up the simulations to managable
     execution times of several months to one year of intensive calculations for about 
     seven years of stellar time.
  A higher numerical resolution shows smaller scale structures appearing within the
granules that are already present in lower resolution simulations \cite[Fig. 10]{2006EAS....18..177C}. 
This, however, should not affect the first few visibility lobes. 

 The approximation of grey radiative transfer is justified only in the stellar interior and 
 it is not appropriate in the optically thin layers. The implementation of non-grey opacities 
 (e.g., five wavelength groups employed to describe the wavelength dependency of the radiation 
 field within a multigroup radiative transfer scheme, see \citealp{1982A&A...107....1N} 
 for details) would be an important improvement for the hydrodynamical simulations.     
\cite{1994A&A...284..105L} found for local RHD simulations that frequency dependent
radiative transfer causes an intensified heat exchange of a fluid element with its 
environment tending to reduce the temperature differences. Consequently, the temperature 
fluctuations in the non-grey local models are smaller than in the grey case. 
This is also expected for global RSG models, and the result of
     such a decrease of the temperature fluctuations, would be a decreased
     intensity contrast and decreased visibility fluctuations. 
 
\begin{acknowledgements}
 This project was supported by the French Ministry of Higher Education through
an ACI (PhD fellowship of Andrea Chiavassa, postdoc fellowship of Bernd Freytag, and computational resources).
Present support is ensured by a grant from ANR (ANR-06-BLAN-0105). 
We are also grateful to the PNPS and CNRS for their financial
support through the years. We thank the CINES for providing some of the computational resources
 necessary for this work.
 We thank Michel Belmas and Philippe Falandry for their help.
Part of this work was made while BPz was on sabbatical at Uppsala Astronomical Observatory.
We thank Bengt Gustafsson, Hans-Gunter Ludwig, John Monnier, Martin Asplund, Nik Piskunov, and Nils Ryde for
enlightening discussions.
\end{acknowledgements}

\bibliographystyle{aa}
\bibliography{biblio.bib}


\end{document}